\documentclass[a4paper,
    12pt,
    abstracton,
    ]{scrartcl}

\usepackage[table]{xcolor}
\usepackage{ifthen}
\usepackage{mathrsfs}
\usepackage{amsmath} %Math commands 
\usepackage{amssymb} %Extra symbols 
\usepackage{amsthm}
\usepackage{color}
\usepackage{hyperref}
\definecolor{refcolor}{RGB}{0,15,255}
\hypersetup{
    colorlinks,
    citecolor=refcolor,
    filecolor=refcolor,
    linkcolor=refcolor,
    urlcolor=refcolor
}
\usepackage{natbib}
\usepackage{multirow}

\usepackage{booktabs}
\usepackage{ifthen}
\usepackage[toc]{multitoc}
\usepackage{tocloft}
\usepackage{setspace}

\newcolumntype{?}{!{\vrule width 1pt}}

\theoremstyle{definition}

\def\({\left(}
\def\){\right)}
\def\[{\left(}
\def\]{\right)}

\newcommand{\tn}{\textnormal}

\newcommand{\mf}[1]{\mathfrak{#1}}
\newcommand{\mc}[1]{\mathcal{#1}}
\newcommand{\ms}[1]{\mathscr{#1}}

\newcommand{\op}[1]{\operatorname{#1}}
\newcommand{\Tr}{\operatorname{Trace}}

\newcommand{\R}{\mathbb{R}}
\newcommand{\N}{\mathbb{N}}
\newcommand{\Z}{\mathbb{Z}}
\newcommand{\C}{\mathbb{C}}
\newcommand{\K}{\mathbb{K}}
\newcommand{\Hq}{\mathbb{H}}
\newcommand{\Oct}{\mathbb{O}}

\newcommand{\gammamod}{\widetilde{\gamma}}
\newcommand{\tgamma}{\Gamma}

\newcommand{\tlambda}{\widetilde{\lambda}}

\newcommand{\Cl}{C\ell}
\newcommand{\CCl}{\C\ell}
\newcommand{\SMA}{\mc{A}_{\op{SM}}}
\newcommand{\D}{\mf{D}}

\newcommand{\GO}{\op{O}}
\newcommand{\SO}{\op{SO}}
\newcommand{\so}{\mf{so}}
\newcommand{\GL}{\op{GL}}

\newcommand{\U}{\op{U}}
\newcommand{\SU}{\op{SU}}
\newcommand{\uu}{\mf{u}}
\newcommand{\su}{\mf{su}}
\newcommand{\End}{\op{End}}

\newcommand{\Spin}{\op{Spin}}
\newcommand{\spin}{\mf{spin}}
\newcommand{\Span}{\op{span}}

\newcommand{\ChargeSpace}[1]{\chi_{\operatorname{#1}}}
\newcommand{\CSEM}{\ChargeSpace{em}}
\newcommand{\CSHyper}{\ChargeSpace{Y}}

\newcommand{\CSColor}{\ChargeSpace{c}}
\newcommand{\CSElectroWeak}{\ChargeSpace{ew}}
\newcommand{\CSElectroColor}{\ChargeSpace{}}

\newcommand{\ChargeAntiSpace}[1]{\overline{\chi}_{\operatorname{#1}}}
\newcommand{\CSAEM}{\ChargeAntiSpace{em}}

\newcommand{\CSAColor}{\ChargeAntiSpace{c}}

\newcommand{\CSAElectroColor}{\ChargeAntiSpace{}}

\newcommand{\SymEM}{\U(1)_{\operatorname{em}}}
\newcommand{\SymHyper}{\U(1)_{\operatorname{Y}}}
\newcommand{\SymWeak}{\SU(2)_{\operatorname{L}}}
\newcommand{\SymElectroWeakOld}{\SymWeak\times\SymHyper}
\newcommand{\SymElectroWeak}{\U(2)_{\operatorname{ew}}}
\newcommand{\SymColor}{\SU(3)_{\operatorname{c}}}
\newcommand{\ssymElectroWeak}{\uu(2)_{\operatorname{ew}}}

\newcommand{\ExtPow}[1]{\bigwedge{}^{#1}}
\newcommand{\ExtAlg}{\ExtPow{\bullet}}

\newcommand{\Matrix}[2]{{\mathbf{M}}_{#2}(#1)}

\newcommand{\qfree}{\mf{q}}
\newcommand{\q}[1]{\qfree_{#1}}
\newcommand{\qd}[1]{\qfree{}\!^\dagger\!{}_{#1}}

\newcommand{\qvol}{\qfree}
\newcommand{\qdvol}{\qvol{}\!^\dagger}
\newcommand{\h}{\mf{h}}

\newcommand{\qqd}{\wproj}
\newcommand{\qdq}{\wproj'}

\newcommand{\e}[1]{\mf{e}_{#1}}
\newcommand{\f}[1]{\tilde{\mf{e}}_{#1}}
\newcommand{\evol}{\e{}}
\newcommand{\fvol}{\f{}}

\newcommand{\bwfree}{\omega}
\newcommand{\bwindex}[1]{
{
  \ifthenelse{\equal{#1}{1}}{\operatorname{u}}{
  \ifthenelse{\equal{#1}{2}}{\operatorname{d}}{
  \ifthenelse{\equal{#1}{3}}{\circ}{#1}}}
}}
\newcommand{\bw}[1]{\bwfree_{\bwindex{#1}}}
\newcommand{\bwd}[1]{\bwfree{}\!^\dagger\!{}_{\bwindex{#1}}}
\newcommand{\bwvol}{\bwfree}
\newcommand{\bwdvol}{\bwvol{}\!^\dagger}

\newcommand{\we}[1]{\mf{u}_{\bwindex{#1}}}
\newcommand{\wf}[1]{\we{#1}'}

\newcommand{\wproj}{\mf{p}}

\newcommand{\wg}[1]{\widetilde{T}_{#1}}

\newcommand{\ns}{\ms{N}}
\newcommand{\nsd}{\ms{N}^\dagger}

\newcommand{\ws}[1]{\mathbb{W}_{#1}}
\newcommand{\wsu}{\ws{0}}
\newcommand{\wsd}{\ws{1}}
\newcommand{\wsuR}{\wsu{}_{\operatorname{R}}}
\newcommand{\wsuL}{\wsu{}_{\operatorname{L}}}
\newcommand{\wsdR}{\wsd{}_{\operatorname{R}}}
\newcommand{\wsdL}{\wsd{}_{\operatorname{L}}}
\newcommand{\wsw}{\ws{w}}

\newcommand{\WeakSpinor}[1]{\psi_{#1}}
\newcommand{\WeakUp}{\WeakSpinor0}
\newcommand{\WeakDown}{\WeakSpinor1}
\newcommand{\WeakUpL}{\WeakUp{}_{\operatorname{L}}}
\newcommand{\WeakUpR}{\WeakUp{}_{\operatorname{R}}}
\newcommand{\WeakDownL}{\WeakDown{}_{\operatorname{L}}}
\newcommand{\WeakDownR}{\WeakDown{}_{\operatorname{R}}}

\newcommand{\hns}{h_{\ns}}
\newcommand{\hw}{h_{\operatorname{w}}}

\newcommand{\eg}{\textit{e.g.}}

\newcommand{\sref}[1]{\S\ref{#1}}

\definecolor{Green}{RGB}{30,210,50}
\definecolor{Yellow}{RGB}{255,220,10}

\newcommand{\pcell}[5]{
\ifthenelse{\equal{#5}{+}}{
\ifthenelse{\equal{#1}{r}}{\cellcolor{red}}{\ifthenelse{\equal{#1}{y}}{\cellcolor{Yellow}}{\ifthenelse{\equal{#1}{b}}{\cellcolor{blue!80}}{\cellcolor{black!80}}}}{\ifthenelse{\equal{#2}{L}}{\color{white!100}{\mathbf{{#3}^{{}_{#1}}_{{}_{#2#4}}}}}{\color{white!100}{\mathbf{{#3}^{{}_{#1}}_{{}_{#2#4}}}}}}
}{
\ifthenelse{\equal{#1}{r}}{\cellcolor{Green}}{\ifthenelse{\equal{#1}{y}}{\cellcolor{violet!80}}{\ifthenelse{\equal{#1}{b}}{\cellcolor{orange}}{\cellcolor{gray!70}}}}{\ifthenelse{\equal{#2}{L}}{\color{white!100}{\mathbf{{\overline{#3}}^{{}_{\overline{#1}}}_{{}_{#2#4}}}}}{\color{white!100}{\mathbf{{\overline{#3}}^{{}_{\overline{#1}}}_{{}_{#2#4}}}}}}
}
}

\begin{document}

\author{Ovidiu Cristinel Stoica
\thanks{Department of Theoretical Physics, National Institute of Physics and Nuclear Engineering -- Horia Hulubei, Bucharest, Romania. Email: \href{mailto:cristi.stoica@theory.nipne.ro}{cristi.stoica@theory.nipne.ro},  \href{mailto:holotronix@gmail.com}{holotronix@gmail.com}}
}
%-----------------------------------------------------%
\title{The Standard Model Algebra}
\subtitle{Leptons, Quarks, and Gauge from the Complex Clifford Algebra \texorpdfstring{$\CCl_6$}{CCl6}}
\maketitle
%\subjclass{Primary 15A66; Secondary 81V22}

%\keywords{Beyond standard model, Grand unified theory, Clifford algebra, Weinberg angle.}

\begin{abstract}
A simple geometric algebra is shown to contain automatically the leptons and quarks of a family of the Standard Model, and the electroweak and color gauge symmetries, without predicting extra particles and symmetries. The algebra is already naturally present in the Standard Model, in two instances of the Clifford algebra $\mathbb{C}\ell_6$, one being algebraically generated by the Dirac algebra and the weak symmetry generators, and the other by a complex three-dimensional representation of the color symmetry, which generates a Witt decomposition which leads to the decomposition of the algebra into ideals representing leptons and quarks.
The two instances being isomorphic, the minimal approach is to identify them, resulting in the model proposed here.
The Dirac and Lorentz algebras appear naturally as subalgebras acting on the ideals representing leptons and quarks. The resulting representations on the ideals are invariant to the electromagnetic and color symmetries, which are generated by the bivectors of the algebra. The electroweak symmetry is also present, and it is already broken by the geometry of the algebra. The model predicts a bare Weinberg angle $\theta_W$ given by $\sin^2\theta_W=0.25$.
The model shares common ideas with previously known models, particularly with Chisholm and Farwell, 1996, Trayling and Baylis, 2004, and Furey, 2016.
\end{abstract}

\setcounter{tocdepth}{1}
\renewcommand{\baselinestretch}{0.75}\normalsize
{\small \tableofcontents}
\renewcommand{\baselinestretch}{1.0}\normalsize

%-----------------------------------------------------%
\section{Introduction}

This article presents an algebra which describes the \emph{Standard Model} (SM) symmetries and a family of leptons and quarks, connecting some of their apparently independent properties like various charges and symmetries in a compact way, whose matrix representation is illustrated in \eqref{eq:standard_model_algebra}.

\begin{equation}
\label{eq:standard_model_algebra}
	\def\arraystretch{1.5}
	\begin{array}{l}
	\\
	\\
	\def\arraystretch{1.5}
	\left.\begin{array}{l}
		\\\tn{Dirac,}
		\\\tn{Lorentz}
		\\
		\\
	\end{array}\right\{
	\\
	\left.\begin{array}{l}
		\\\tn{Dirac,}
		\\\tn{Lorentz}
		\\
		\\
	\end{array}\right\{
	\\
	\end{array}
	%%%
	\def\arraystretch{1.5}
	\begin{array}{l}
	\\
	\\
	\def\arraystretch{3.0}
	\left.\begin{array}{l}
		\mathbf{1}_w
	\end{array}\right\{
	\\
	\def\arraystretch{2.0}
	\left.\begin{array}{l}
		\\ \mathbf{2}_w
		\\
		\\
	\end{array}\right\{
	\\
	\def\arraystretch{3.0}
	\left.\begin{array}{l}
		\mathbf{1}_w
	\end{array}\right\{
	\end{array}
	%%%
\arraycolsep=1.5pt\def\arraystretch{1.5}
\begin{array}{llllllll}
\multicolumn{1}{c}{\mathbf{1}_c} & \multicolumn{3}{c}{\mathbf{3}_c} & \multicolumn{1}{c}{\overline{\mathbf{1}}_c} & \multicolumn{3}{c}{\overline{\mathbf{3}}_c} \\
\overbrace{}^{} & \multicolumn{3}{c}{\overbrace{\hspace{55pt}}^{}} & \overbrace{}^{} & \multicolumn{3}{c}{\overbrace{\hspace{55pt}}^{}} \\
		\pcell {}R\nu1+ & \pcell rRu1+ & \pcell yRu1+ & \pcell bRu1+ & \pcell {}Le1- & \pcell rLd1- & \pcell yLd1- & \pcell bLd1- \\
		\pcell {}R\nu2+ & \pcell rRu2+ & \pcell yRu2+ & \pcell bRu2+ & \pcell {}Le2- & \pcell rLd2- & \pcell yLd2- & \pcell bLd2- \\
		\\[-21pt]
		\pcell {}L\nu1+ & \pcell rLu1+ & \pcell yLu1+ & \pcell bLu1+ & \pcell {}Re1- & \pcell rRd1- & \pcell yRd1- & \pcell bRd1- \\
		\pcell {}L\nu2+ & \pcell rLu2+ & \pcell yLu2+ & \pcell bLu2+ & \pcell {}Re2- & \pcell rRd2- & \pcell yRd2- & \pcell bRd2- \\
		\\[-20.5pt]
		\pcell {}Le1+ & \pcell rLd1+ & \pcell yLd1+ & \pcell bLd1+ & \pcell {}R\nu1- & \pcell rRu1- & \pcell yRu1- & \pcell bRu1- \\
		\pcell {}Le2+ & \pcell rLd2+ & \pcell yLd2+ & \pcell bLd2+ & \pcell {}R\nu2- & \pcell rRu2- & \pcell yRu2- & \pcell bRu2- \\
		\\[-21pt]
		\pcell {}Re1+ & \pcell rRd1+ & \pcell yRd1+ & \pcell bRd1+ & \pcell {}L\nu1- & \pcell rLu1- & \pcell yLu1- & \pcell bLu1- \\
		\pcell {}Re2+ & \pcell rRd2+ & \pcell yRd2+ & \pcell bRd2+ & \pcell {}L\nu2- & \pcell rLu2- & \pcell yLu2- & \pcell bLu2- \\
	\end{array}
\end{equation}

A three-dimensional Hermitian space $\CSElectroColor$ determines a Clifford algebra $\CCl\(\CSElectroColor^\dagger \oplus \CSElectroColor\)$, which is naturally split into left ideals.
In a basis adapted to the ideal decomposition, each column contains two $4$-spinors associated to different flavors, as illustrated in \eqref{eq:standard_model_algebra}.
The Lie group $\SymColor$ permutes the columns according to the representations $\mathbf{1}_c$, $\mathbf{3}_c$, $\overline{\mathbf{1}}_c$, and $\overline{\mathbf{3}}_c$.
Each ideal is indexed with an electric charge which is multiple of $\frac13$ partially representing the charge of the upper particle, and its color is determined by the ideal to which belongs, having associated a particular representation of $\SymColor$.
The actions of the Dirac algebra and the Lorentz group are reducible, permute the rows of each ideal, and split it naturally into two $4$-spinors, whose left chiral components are permuted by the $\SymWeak$ symmetry by the representations $\mathbf{1}_w$ and $\mathbf{2}_w$.
Hence, the leptons, quarks, and gauge symmetries of the SM are reproduced properly.

%%%%%%%%%
In the $\SU(5)$ and $\Spin(10)$ unified theories the basis of the exterior algebra $\ExtAlg\(\C^2\oplus\C^3\)$ classifies the internal degrees of freedom of leptons and quarks, the Clifford algebra $\Cl_{10}$ being generated by ladder operators acting on $\ExtAlg\(\C^2\oplus\C^3\)$ ({\eg} \cite{BaezHuerta2010GUTAlgebra}).
The model proposed here uses the representation of colors and electric charges from $\ExtAlg\C^3$, realized as a minimal right ideal of $\CCl_6$, which classifies the minimal left ideals by a standard construction using the Witt decomposition \citep{chevalley1997algebraicspinors,crumeyrolle1990clifford}. This has the desired symmetries, and leptons and quarks are represented on each of the resulting minimal right ideals.

The model proposed here shares common features with previously known models.
%%% idempotents: Chisholm and Farwell, 1996, Trayling and Baylis, 2004
Particles of two distinct flavors were previously combined into $8$-spinor ideals,  in a unified spin gauge theory of gravity and electroweak interactions based on $\Cl_{1,6}\cong\Cl_{1,3}\otimes\Cl_{0,3}$ \citep{ChisholmFarwell1996CliffordAlgebrasForFundamentalParticles},
and in \citep{Trayling1999GeometricApproachStandardModel,TraylingBaylis2001GeometricStandardModelGaugeGroup,TraylingBaylis2004Cl7StandardModel} based on $\Cl_7\cong\Cl_3\otimes\Cl_4$, where there are three space dimensions, the time is a scalar, the four extra dimensions related to the Higgs boson, the predicted Weinberg angle is given by $\sin^2\theta_W=0.375$, and remarkably, the full symmetries of the SM arise from the condition to preserve the current and to leave right-handed neutrino sterile.
Among the main differences, the model proposed here uses different structures, leading to the algebra $\CCl(\CSElectroColor^\dagger \oplus \CSElectroColor)\cong\CCl_6$, includes the Dirac algebra $\Cl_{1,3}\otimes\C$, and $\sin^2\theta_W=0.25$.
In the $\Cl_{1,6}$ and $\Cl_7$ models the ideals are obtained using primitive idempotents.
The model proposed here uses a decomposition into left ideals $\CCl_6\qvol\qdvol\q K$, where $\K\subseteq\{1,2,3\}$ (notations from \sref{s:sma_ideal}), based on the Witt decomposition $\CSElectroColor^\dagger \oplus \CSElectroColor$ and the exterior algebra $\ExtPow k\CSElectroColor$ contained within the minimal right ideal $\qvol\qdvol\CCl_6=\qvol\qdvol\ExtPow k\CSElectroColor$.
This allows the colors and charges to be emphasized, and the minimal left ideals of the same charge and different colors to be coupled into a larger ideal.
%%% Furey, 2016
In a model based on octonions, Furey \citep{Furey2015ChargeQuantization,Furey2016StandardModelFromAlgebra} uses the Witt decomposition for $\CCl_6$ to represent colors and charges of up- and down-type particles by $\qd K\qvol\qdvol$ and $\q K\qdvol\qvol$, on the minimal left ideals $\CCl_6\qvol\qdvol$ and $\CCl_6\qdvol\qvol$.
They are united into a single irreducible representation of $\CCl_6\otimes_\C\CCl_2$ obtained by using the octonion algebra.
To represent the complete particles, with spin and chirality, Furey proposes including the quaternion algebra,
resulting in a representation of leptons and quarks as spinors of an algebra isomorphic to $\CCl_{12}$.
By contrast, in the model proposed here, everything is contained in the ideals of $\CCl_6$ classified by the elements $\qvol\qdvol\q K$.
Despite these differences, the $\SymColor$ and $\SymEM$ symmetries in the model proposed here are identical to those obtained previously by Furey \citep{Furey2016StandardModelFromAlgebra} as the unitary spin transformations preserving the Witt decomposition of $\CCl_6$, improving by this previous results based on octonions and Clifford algebras \citep{GunaydinGursey1974QuarkStatisticsOctonions,Barducci1977QuantizedGrassmannVariablesGUT,Casalbuoni1979UnifiedQuarksLeptons}.
Note that the algebra used by Furey is the Dixon algebra $\C\otimes\Hq\otimes\Oct$. Dixon used previously this algebra to build a model for leptons and quarks in a model of high mathematical beauty and important physical insights \citep{Dixon2013DivisionAlgebras}.
A unified model based on the Clifford algebra $\Cl_{1,5}$, which predicts the same Weinberg angle of $30^\circ$, was proposed in \citep{Besprosvany2000GaugeSpaceTimeUnification}.
Another model, based on $\Cl^\ast_3\cong\GL(2,\C)$ for the electroweak unification and the lepton sector, and on the Clifford algebra $\Cl_{1,5}$ to include quarks and obtain the full unification with $\SymColor$, was developed in \citep{Daviau2017StandardModelClifford15,Daviau2015StandardModelCliffordAlgebra}. The authors obtained the same Weinberg angle of $30^\circ$ in a model based on de Broglie's idea on the photon as made of two Dirac waves \citep{Daviau2015RetourALOndeDeLouisDeBroglie,Daviau2015WeinbergAngle}.
An interesting approach to grand unification, based on the exceptional real Jordan algebra of dimension $27$, was proposed in \citep{DuboisViolette2016JordanAlgebraGUT,TodorovDuboisViolette2017JordanAlgebraGUT}.
A unification of conformal gravity with an extended standard model with four families was proposed in \citep{Castro2017CliffordUnificationConformalGravityExtendedStandardModel}, based on the Clifford algebra $\CCl_5$.
In \citep{Vargas2013U1xSU2TangentBundle} a way to obtain $\SymHyper\times\SymWeak$ from the tangent bundle was proposed.
Another approach, based on the Clifford algebra $\Cl_{6,0}$ obtained from the non-relativistic phase space, has been proposed in \citep{Zenczykowski2017CliffordAlgebraStandardModel,Zenczykowski2013ElementaryParticlesEmergentPhaseSpace}.
There are significant differences between all these models, and they all are different from the one proposed here, but it would take a dedicated review to describe all, so I just compared here the ones with the most relevant common parts with the model proposed here.

In the following, starting from the algebra $\SMA:=\CCl(\CSElectroColor^\dagger \oplus \CSElectroColor)$ alone, we will recover the properties of leptons, quarks, and the symmetries of the SM.
I was basically forced to arrive at the complex Clifford algebra $\CCl_6$ by the need to include the Dirac algebra and the SM forces.
On the one hand, the Dirac algebra combined with the weak interaction between two particles generate a Clifford algebra $\CCl_6$, which therefore is implicit in the SM (Section \sref{s:from_weak_to_ccl_six}).
On the other hand, the representations of the $\SymColor$ group in the SM and the fact that the electric charge is multiple of $\frac13$ lead to the exterior algebra of a complex three-dimensional vector space $\CSElectroColor$, and its dual $\CSElectroColor^\dagger$ (Section \sref{s:patternsSM}), which together generate the Clifford algebra $\CCl(\CSElectroColor^\dagger \oplus \CSElectroColor)\cong\CCl_6$ (Section \sref{s:sma}).
Thus, any structure aiming to describe the particles and forces of the SM has to include two instances of the algebra $\CCl_6$. Since the two instances are isomorphic, the minimal solution is to identify them. This minimal algebraic structure is the algebra $\SMA$.
This model does not accommodate at this moment more families, and it does not discuss masses, various mixing angles and phases, and for simplicity I will denote the leptons and quarks by the names of the first family.

%-----------------------------------------------------%
\section{Motivation}

In the following, the \emph{Dirac algebra} $\D:=\Cl_{1,3}\otimes\C\cong\CCl_{4}$ is the complexification of the \emph{real Clifford algebra} $\Cl_{1,3}$ associated to the \emph{Lorentz metric} having the signature $(+,-,-,-)$.
The Dirac algebra $\CCl_{4}$ is isomorphic to the complex matrix algebra $\Matrix\C4$, or equivalently, to the algebra of complex linear transformations $\End_\C(W)$ of a four-dimensional complex space $W$.

The \emph{Standard Model} of particle physics, extended with the data about neutrinos, contains all we know about elementary particles.
The fundamental fermions are the \emph{leptons} and \emph{quarks}, and their antiparticles.

Three charged leptons are known: the \emph{electron} $e^-$, the \emph{muon} $\mu^-$, and the \emph{taon} $\tau^-$, having as antiparticles respectively the \emph{positron} $e^+$, the antimuon $\mu^+$, and the antitaon $\tau^+$.
We also know so far three neutral leptons, named \emph{neutrinos}: the \emph{electron neutrino} $\nu_{e}$, the \emph{muon neutrino} $\nu_{\mu}$, and the \emph{tau neutrino} $\nu_{\tau}$. Their corresponding antiparticles are respectively the \emph{electron antineutrino} $\overline{\nu}_{e}$, the \emph{muon antineutrino} $\overline{\nu}_{\mu}$, and the \emph{tau antineutrino} $\overline{\nu}_{\tau}$.

The fundamental fermions of a generic family and their discrete properties are summarized in Table \ref{tab:standard_model_fermions}.

\def\arraystretch{1.5}%  1 is the default, change whatever you need
\begin{table}[ht]
\begin{center}
    \begin{tabular}{ ? l l ? r | r | r | r ? r | r | r | r ? p{3cm} |}
    \bottomrule[1.5pt]\rowcolor{gray!50}
    \textbf{Particle} & & $\mathbf{e^-}$ & $\mathbf{\overline u}$ & $\mathbf{d}$ & $\mathbf{\overline \nu}$ & $\mathbf{\nu}$ & $\mathbf{\overline d}$ & $\mathbf{u}$ & $\mathbf{e^+}$ \\\hline
    \textbf{Electric charge} & & $-1$ & $-\frac{2}{3}$ & $-\frac{1}{3}$ & $0$ & $0$ & $+\frac{1}{3}$ & $+\frac{2}{3}$ & $+1$ \\ \hline
		\multirow{2}{*}{\textbf{Weak isospin}}
		& \textbf{L} & $-\frac{1}{2}$ & $0$ & $-\frac{1}{2}$ & $0$ & $+\frac{1}{2}$ & $0$ & $+\frac{1}{2}$ & $0$ \\
		& \textbf{R} & $0$ & $-\frac{1}{2}$ & $0$ & $-\frac{1}{2}$ & $0$ & $+\frac{1}{2}$ & $0$ & $+\frac{1}{2}$ \\ \hline
		\multirow{2}{*}{\textbf{Hypercharge}}
		& \textbf{L} & $-1$ & $-\frac{4}{3}$ & $+\frac{1}{3}$ & $0$ & $-1$ & $+\frac{2}{3}$ & $+\frac{1}{3}$ & $+2$ \\
		& \textbf{R} & $-2$ & $-\frac{1}{3}$ & $-\frac{2}{3}$ & $1$ & $0$ & $-\frac{1}{3}$ & $+\frac{4}{3}$ & $+1$ \\
		\toprule[1.25pt]
    \end{tabular}
\end{center}
\caption{Discrete properties of leptons and quarks.}
\label{tab:standard_model_fermions}
\end{table}

As we know from Wigner \citep{Wig39}, particles are classified by unitary representations of the Poincar\'e group. Because the Poincar\'e group is noncompact, the unitary representations can only be infinite-dimensional. Particles are classified by spin and mass. Here we will not deal with the mass, since we will focus on the values of field at each point, which admits finite-dimensional representations. So the question ``if all possible combinations of mass and spin are allowed by Wigner's theorem, why only some of them correspond to real particles?'' will not be addressed here.

In addition to spin and mass, particles have \emph{internal degrees of freedom}.
The \emph{gauge symmetry} is the symmetry of an internal space. The Standard Model contains as internal symmetry groups the \emph{color group} $\SymColor$, and the \emph{electroweak group}, which contains the \emph{electromagnetic group} $\SymEM$. It is usually considered that the electroweak group is $\SymElectroWeakOld$. What we really know is not the Lie group, but its Lie algebra. There are strong arguments that the more natural, both geometrically and physically, Lie group of the electroweak symmetry is not $\SymElectroWeakOld$, but $\SymElectroWeak$, which is locally isomorphic to it,  $\SymElectroWeak\cong\SymElectroWeakOld/\Z_2$ (\cite{Derdzinski1992GeometryStandardModel}, Ch. 6), (\cite{McCabe2011StructureInterpretationStandardModel}, Ch. 4). The results presented here also support this position, so I will take $\SymElectroWeak$ as the electroweak group.

In the following I will ignore the families and the masses, and I will consider as prototype the first family.
I will try to construct the simplest model that answers the following questions:
\begin{enumerate}
	\item 
Why each family contains a charged and a neutral lepton, and two quarks?
	\item 
Why are leptons white, that is, why they do not have color charges?
	\item 
Why are the charges positive or negative multiples of $\frac{1}{3}$ of the electron's charge?
Why the charge of the elementary fermions is never larger, in absolute value, than the charge of the electron?
	\item 
Why are the charges of the leptons and quarks how they are, that is, why for example there is no down quark with positive charge, or other combination?
	\item 
Why does the Standard Model's internal symmetry group is particularly $\SymColor\times\SymElectroWeak$, when infinitely many other possibilities are available?
	\item 
Why does the Standard Model apparently contains only fundamental particles corresponding to some of the possible representations of these gauge groups, when infinitely many other possibilities are available?
	\item 
Why does the electroweak force depend on the chirality of the particle, which is related to spin?
\end{enumerate}

Some of these questions are addressed and answered by \emph{grand unified theories} (GUT) like $\SU(5)$ \citep{georgi1974GUTSU5} and $\SO(10)$ (actually $\Spin(10)$) \citep{georgi1975GUTSO10,HaraldMinkowski1975GUTSO10}. So far all such models predict still undetected additional interactions and proton decay. But they contain important insights for future developments.

After the success of the electroweak unification, the natural question was ``is it possible to unify the electroweak symmetry with the color symmetry?''. This suggested that a larger group, which extends the Standard Model group but is simple, should exist. The requirement that the group should be simple comes from the desire to unify the coupling constants.
Geometrically speaking, the coupling constants, which appear as coefficients in the Lagrangian, are coefficients defining on the Lie algebra an inner product invariant to the adjoint action of the Lie group on the Lie algebra.
If the unifying Lie group is simple, it admits a unique, up to a constant factor, Ad-invariant inner product. The coupling constants of the forces that are unified are thus obtained as coefficients of the inner products induced by the unification inner product on their Lie algebras viewed as Lie subalgebras of the unifying Lie algebra.
These resulted in the $\SU(5)$ and $\Spin(10)$ GUTs. But the coupling constants are in fact properties of the group's Lie algebra. So another possibility is that the Standard Model is unified in a simple algebra, which is not necessarily a Lie algebra, but it has a naturally associated Lie algebra. There are two problems with unification theories based on simple groups. First, they invariably lead to the need of another symmetry breaking, additional gauge bosons, and proton decay. Second, simply finding a unifying group does not explain the representations chosen by Nature. A Lie group has infinitely many representations, so we would need an additional principle to pick the right representation. In the case of the $\SU(5)$ GUT the representation is $\overline{\mathbf{5}} \oplus \mathbf{10}$ (and optionally a singlet $\mathbf{1}$ containing the right-handed neutrino), while in the $\Spin(10)$ GUT, which extends $\SU(5)$, the representation reduces to $\overline{\mathbf{5}} \oplus \mathbf{10} \oplus \mathbf{1}$. The only justification for these representations comes from the compatibility with the known particles in the Standard Model, but the question ``why this one, among infinitely many possible representations?'' remains.

But what if there is a simple mathematical structure which has the symmetries of the Standard Model, and provides the right representations automatically? In the following I will propose that the complex Clifford algebra $\CCl_6$ with a certain ideal decomposition (denoted here by $\SMA$) has the desired features.
The Dirac algebra arises as a distinguished subalgebra of $\SMA$. The group of transformations of $\SMA$ compatible with the ideal decomposition and the Dirac algebra is just the Standard Model group.

Why the complex Clifford algebra $\CCl_6$? In Section \sref{s:from_weak_to_ccl_six} I show that $\CCl_6$ is inherent in the Standard Model: just take the internal space of a pair of weakly interacting particles, such as $\nu_{e}$ and $e^-$. The weak force generators act on the direct sum of their spinor spaces, which is an eight-dimensional complex vector space, and extend the Dirac algebra to the algebra $\CCl_6$.
The Clifford algebra $\CCl_6$ also results naturally if we inspect the representations of $\SymEM$ and $\SymColor$ found in the SM (Sections \sref{s:patternsSM}, \sref{s:sma}). The minimal model containing both instances of $\CCl_6$ is the one in which they are identified, and which is presented here.
The main part of the article is devoted to explaining how this algebra accounts automatically for the quarks and the symmetries associated with the known interactions, at least for a family.

%-----------------------------------------------------%
\section{Patterns in the Standard Model}
\label{s:patternsSM}

First, let us organize a little bit the Standard Model.
The internal charge spaces are summarized in Table \ref{tab:standard_model_charge_spaces}:
\def\arraystretch{1.5}
\begin{table}[ht]
\begin{center}
    \begin{tabular}{ ? l ? c | c | c ? p{2.5cm} |}
    \bottomrule[1.5pt]\rowcolor{gray!50}
    \textbf{Type of charge / force} & \textbf{Internal space} & \textbf{$\C$-dimension} & \textbf{Symmetry} \\\hline
    \textbf{Electromagnetic} & $\CSEM$ & $1$ & $\SymEM$ \\ \hline
    \textbf{Weak hypercharge} & $\CSHyper$ & $1$ & $\SymHyper$ \\ \hline
    \textbf{Color} & $\CSColor$ & $3$ & $\SymColor$ \\ \hline
    \textbf{Electroweak} & $\CSElectroWeak$ & $2$ & $\SymElectroWeak$ \\
    \toprule[1.5pt]
    \end{tabular}
\end{center}
\caption{The internal charge spaces of particles in the Standard Model.}
\label{tab:standard_model_charge_spaces}
\end{table}

In the following, if $(V,h)$ is a complex vector space $V$ with a Hermitian inner product $h$, $h$ establishes canonical complex linear isomorphisms between the dual and the conjugate of $V$, $V^\dagger\cong\overline{V}$, and also $V\cong\overline{V}^\dagger$. We will also use the notation $V^{-1}:=V^\dagger=\overline{V}$, which comes handy when talking about charges.

Let us consider a number of $n$ particles of electric charges $Q_i$, $i\in\{1,\ldots,n\}$. The total system of these particles is represented as a vector in the tensor product of the state vector spaces of the individual particles, and its charge is $Q=\sum_{i=1}^n Q_i$. This applies to tensor and exterior products of internal states too.

In particular, the one-dimensional complex space $\CSEM$ is endowed with a Hermitian inner product $h_{em}$, and is the internal elementary charge space, representing a charge equal to $\frac{1}{3}\mathit{e}$, where $\mathit{e}$ is the electron charge.
The charge $-\frac{1}{3}\mathit{e}$ is represented on the complex conjugate space $\CSAEM=\CSEM^{-1}$.
For a particle of charge $\frac{k}{3}\mathit{e}$, where $k\in\Z$, the internal charge space is $\CSEM^k:=\bigotimes^k_{\C}\CSEM$ \citep{PeR87,Derdzinski1992GeometryStandardModel}.

The combined internal charge and color space for leptons and quarks are represented in Table \ref{tab:electric_color_charge_spaces}.
\def\arraystretch{1.5}
\begin{table}[ht]
\begin{center}
    \begin{tabular}{ ? l ? c | c | c | c ? c | c | c | c ? p{3cm} |}
    \bottomrule[1.5pt]\rowcolor{gray!50}
    \textbf{Particle} & $\mathbf{e^-}$ & $\mathbf{\overline u}$ & $\mathbf{d}$ & $\mathbf{\overline \nu}$ & $\mathbf{\nu}$ & $\mathbf{\overline d}$ & $\mathbf{u}$ & $\mathbf{e^+}$ \\\hline
    \textbf{Electro-color space} & $\CSEM^3$ & $\CSEM^2\CSAColor$ & $\CSEM^1\CSColor$ & $\C$ & $\C$ & $\CSEM^{-1}\CSAColor$ & $\CSEM^{-2}\CSColor$ & $\CSEM^{-3}$ \\
    \toprule[1.5pt]
    \end{tabular}
\caption{The internal charges spaces of particles.}
\label{tab:electric_color_charge_spaces}
\end{center}
\end{table}

Taking tensor products of the color space $\CSColor$ with itself would allow various internal combinations of colors, for example multiples of $r$, $y$, and $b$, which seem to be absent in nature.
Why are there only (white) leptons, and quarks of colors $r$, $y$, and $b$, and antiquarks of colors $\overline{r}$, $\overline{y}$, and $\overline{b}$?
A possible answer is that the only color spaces are exterior powers of $\CSColor$, that is, $\ExtPow k\CSColor$ and $\ExtPow {-k}\CSColor:=\ExtPow k\CSAColor$, where $k\in\{0,1,2,3\}$.
Then, the only possibilities would be $\ExtPow 0\CSColor\cong\C$, $\ExtPow 1\CSColor\cong\CSColor$, $\ExtPow 2\CSColor$, $\ExtPow 3\CSColor\cong\C$, and their complex conjugate spaces. This reproduces the color and charge patterns of leptons, quarks, and their antiparticles.

But there is an isomorphism between $\ExtPow 2\CSColor$ and $\CSAColor$, obtained as follows. Consider a volume form on $\CSColor$, that is, a unit vector $\lambda_c\in\ExtPow 3\CSColor^\dagger$. Then, for any $\xi\in\ExtPow 2\CSColor$, the contraction $\lambda_c^{ABC}\omega_{AB}$ is a vector from $\CSColor^\dagger\cong\CSAColor$. This isomorphism is unique up to a phase factor, and it acts by $r\wedge y \mapsto \overline{b}$, $y\wedge b \mapsto \overline{r}$, $b\wedge r \mapsto \overline{y}$, where $(r,y,b)$ is a basis in $\CSColor$.

Let us define $\CSElectroColor:=\CSAEM\otimes\CSAColor$ (the justification of this choice rather than $\CSEM\otimes\CSColor$ will follow). The space $\CSElectroColor$ has complex dimension three and is Hermitian, with the Hermitian inner product $\h=h_{em}\otimes h_c$. 

Let us pick an orthonormal (with respect to the Hermitian inner product $\h$) basis of $\CSElectroColor$, and its dual basis in $\CSElectroColor^\dagger$, as
\begin{equation}
\label{eq:basis_CSElectroColor}
\begin{cases}
(\q 1,\q 2,\q 3) \\
(\qd 1,\qd 2,\qd 3). \\
\end{cases}
\end{equation}
The inner product $\h$ induces a positive definite Hermitian inner product on $\CSAElectroColor$, with respect to which the basis $(\qd 1,\qd 2,\qd 3)$ is orthonormal.

Let us choose the following basis for $\ExtAlg\CSElectroColor$,
\begin{equation}
\label{eq:basis_CSElectroColorExt}
(1, \q{23}, \q{31}, \q{12}, \q{321}, \q{1}, \q{2}, \q{3}),
\end{equation}
where $\q{j_1\ldots j_k} := \q{j_1}\ldots \q{j_k}$.
The inner product $\h$ extends on the exterior algebra $\ExtAlg\CSElectroColor$ to a positive definite Hermitian inner product for which the basis \eqref{eq:basis_CSElectroColorExt} is orthonormal.

The combined internal charge and color spaces for leptons and quarks are represented now in Table \ref{tab:electric_color_charge_spaces_unified}.

\def\arraystretch{1.5}
\begin{table}[ht]
\begin{center}
    \begin{tabular}{ ? l ? c | c | c | c ? c | c | c | c ? p{3cm} |}
    \bottomrule[1.5pt]\rowcolor{gray!50}
    \textbf{Particle} & $\mathbf{e^-}$ & $\mathbf{\overline u}$ & $\mathbf{d}$ & $\mathbf{\overline \nu}$ & $\mathbf{\nu}$ & $\mathbf{\overline d}$ & $\mathbf{u}$ & $\mathbf{e^+}$ \\\hline
    \textbf{Electro-color space} & $\ExtPow 3\CSAElectroColor$ & $\ExtPow 2\CSAElectroColor$ & $\ExtPow 1\CSAElectroColor$ & $\ExtPow 0\CSAElectroColor$ & $\ExtPow 0\CSElectroColor$ & $\ExtPow 1\CSElectroColor$ & $\ExtPow 2\CSElectroColor$ & $\ExtPow 3\CSElectroColor$ \\
    \toprule[1.5pt]
    \end{tabular}
\caption{The unified internal charges spaces, where $\CSElectroColor:=\CSAEM\CSAColor$.}
\label{tab:electric_color_charge_spaces_unified}
\end{center}
\end{table}

Now all leptons and quarks, and their antiparticles, are classified by exterior powers of $\CSElectroColor$, and some of the relations between various discrete parameters are explained, as well as the Standard Model gauge group after the symmetry breaking. 
The spaces $\ExtPow k\CSElectroColor$ are the desired representations of both $\SymEM$ and $\SymColor$.

As for the weak interaction between two particles, for example $\nu_{e}$ and $e^-$, their spinor spaces decompose in the singlet spaces for $\nu_{eR}$ and $e^-_R$, and a doublet space for $\nu_{eL}$ and $e^-_L$.
These spaces also correspond to the exterior powers of a certain complex vector space, as we shall see in Section \sref{s:weak_symmetry_sma}.

Note that GUTs like $\SO(5)$ and $\Spin(10)$ are based on representations on the exterior algebra of a $5$-dimensional complex space, $\ExtAlg \C^5$, which includes a subspace similar (though not identical) to $\ExtAlg\CSElectroColor$, and another one corresponding to the electroweak interaction.
These subspaces are obtained by a symmetry breaking -- in addition to the electroweak symmetry breaking -- which predicts other $12$ gauge bosons that exchange leptons and quarks.
I am interested here to avoid, if possible, this additional symmetry breaking and the resulting bosons and proton decay, which seem to be ruled out by experiments.

In the following, the exterior algebra corresponding to the electric and color interactions, and that corresponding to the weak interaction, will turn out to be subalgebras of the Clifford algebra $\CCl_6$. Leptons and quarks, with their charges and interactions, will emerge out of the same algebra.

%-----------------------------------------------------%
\section{An algebra for the Standard Model}
\label{s:sma}

In this Section I construct an algebra which will turn out to contain the symmetries and generic family of the Standard Model, with their properties.
The method of construction of a Clifford algebra out of a vector space and its dual was employed for example in \citep{DoranHestenes1993LGasSG,Gualtieri2004GeneralizedComplexGeometryThesis}.
This algebra turns out to be the Clifford algebra $\CCl_6$ endowed with a Witt decomposition, which leads to an ideal decomposition classified by the exterior powers of $\CSElectroColor$, which in turn can be used to classify particles by the $\SU(3)$ symmetries.

%-----------------------------------------------------%
\subsection{Definition and main properties}
\label{s:sma_def}

On the space $\CSElectroColor^\dagger \oplus \CSElectroColor$ we define the \emph{inner product}
\begin{equation}
\label{eq:sma_inner_prod}
\langle u_1^\dagger + u_2, u_3^\dagger + u_4 \rangle := \dfrac12\(u_1^\dagger(u_4) + u_3^\dagger(u_2)\) \in \C,
\end{equation}
where $u_1^\dagger,u_3^\dagger\in\CSElectroColor^\dagger$ and $u_2,u_4\in\CSElectroColor$ (also see \cite{Gualtieri2004GeneralizedComplexGeometryThesis}). This inner product should not be confounded with the Hermitian inner product $\h$ on $\CSElectroColor$. In fact, $\CSElectroColor$ and $\CSElectroColor^\dagger$ are maximally isotropic subspaces for the inner product \eqref{eq:sma_inner_prod}.

We denote by $\SMA$ the Clifford algebra defined by the inner product \eqref{eq:sma_inner_prod},
\begin{equation}
\label{eq:sma}
\SMA := \CCl(\CSElectroColor^\dagger \oplus \CSElectroColor)\cong \CCl_6,
\end{equation}
together with the \emph{Witt decomposition} $\CSElectroColor^\dagger \oplus \CSElectroColor$ of the base complex $6$-dimensional space, and with the Hermitian inner product on $\CSElectroColor$ and $\CSElectroColor^\dagger$.

The elements of the bases defined in equation \eqref{eq:basis_CSElectroColor} satisfy the anticommutation relations
\begin{equation}
\label{eq:sma_anticommutation_relations}
\begin{array}{l}
\{\q j, \q k\} = 0, \\
\{\qd j, \qd k\} = 0, \\
\{\q j, \qd k\} = \delta_{jk} \\
\end{array}
\end{equation}
for $j,k\in\{1,2,3\}$ (also see \cite{BaezHuerta2010GUTAlgebra,Furey2016StandardModelFromAlgebra}).
In other words, $(\qd 1,\qd 2,\qd 3,\q 1,\q 2,\q 3)$ is a \emph{Witt basis} of $\CSElectroColor^\dagger \oplus \CSElectroColor$.

The map $\q j \mapsto \qd j$, $j\in\{1,2,3\}$ corresponds to the \emph{main antiautomorphism} of the Clifford algebra $\SMA$, $\dagger:\SMA\to\SMA$, defined by
\begin{equation}
\label{eq:w_dagger}
\begin{cases}
c^\dagger = c^\ast,\\
\(\q j\)^\dagger=\qd j, \(\qd j\)^\dagger=\q j,\\
(a+b)^\dagger = a^\dagger + b^\dagger,\\
(ab)^\dagger = b^\dagger a^\dagger,\\
\end{cases}
\end{equation}
for any $a,b\in\SMA$, $c\in\SMA^0=\C$.
Then, $(a^\dagger)^\dagger=a$ for any $a\in\SMA$. As we shall see, the algebra $\SMA$ can be represented as a matrix algebra, such that $\dagger$ corresponds to the matrix adjoint.

We define an orthonormal basis of the vector space $\CSElectroColor^\dagger \oplus \CSElectroColor$, $(\e 1,\e 2,\e 3,\f 1,\f 2,\f 3)$, by
\begin{equation}
\label{eq:basis_sma}
\begin{cases}
\e j = \q j + \qd j \\
\f j = i\(\qd j - \q j\), \\
\end{cases}
\end{equation}
where $j\in\{1,2,3\}$.
Then, $\e j^2=1$, $\f j^2=1$, $\e j^\dagger=\e j$, and $\f j^\dagger=\f j$. Also,
\begin{equation}
\label{eq:basis_sma_back}
\begin{cases}
\q j = \frac12\(\e j + i\f j\) \\
\qd j = \frac12\(\e j - i\f j\). \\
\end{cases}
\end{equation}

The algebra $\SMA$ inherits the gradation from its Clifford algebra structure, $\SMA=\bigoplus_{k=0}^{6}\SMA^k$, where the vector subspaces $\SMA^k$ of pure degree $k$ correspond to those of the exterior algebra $\ExtAlg\(\CSElectroColor^\dagger \oplus \CSElectroColor\)$,
\begin{equation}
\label{eq:sma_clifford_ext}
\SMA^k=\CCl^k(\CSElectroColor^\dagger \oplus \CSElectroColor) = \ExtPow k\(\CSElectroColor^\dagger \oplus \CSElectroColor\).
\end{equation}

We define the elements
\begin{equation}
\label{eq:efvol}
\begin{cases}
\evol :=& \e 1 \e 2 \e 3, \\
\fvol =& \f 1 \f 2 \f 3. \\
\end{cases}
\end{equation}
Then, $\evol^2=-1$, $\fvol^2=-1$, $\evol\fvol=-\fvol\evol$, and $(\evol\fvol)^2=-1$.

As for any Clifford algebra, we define the \emph{main automorphism} of $\SMA$ as $\alpha:\SMA\to\SMA$,
\begin{equation}
\label{eq:sma_clifford_main_automorphism}
\begin{cases}
\alpha(a) := -a & \tn{ for any } a\in\SMA^1,\\
\alpha(ab) := \alpha(a)\alpha(b) & \tn{ for any } a,b\in\SMA.
\end{cases}
\end{equation}
It is easy to see that for any $a\in\SMA^k$, $\alpha(a)=(-1)^k a$.

The operation $a\mapsto\tilde{a}$ extends to $\SMA$ by $\tilde{\tilde{a}}=a$ and $\widetilde{ab}=\tilde{b}\tilde{a}$ for any $a,b\in\SMA$.

%-----------------------------------------------------%
\subsection{Ideals and representation}
\label{s:sma_ideal}

The elements
\begin{equation}
\label{eq:qvol}
\begin{cases}
\qvol :=& \q 1 \q 2 \q 3,\\
\qdvol =& \qd 3 \qd 2 \qd 1, \\
\end{cases}
\end{equation}
are \emph{nilpotent}, since $\qvol^2=0$ and $\qdvol{}^2=0$.

Let us make the notation $\qqd:=\qvol\qdvol$ and $\qdq=\qdvol\qvol$, since we will use them extensively.
It can be shown that
\begin{equation}
\wproj  = \frac{1 + i \e 1 \f 1}{2} \cdot \frac{1 + i \e 2 \f 2}{2} \cdot \frac{1 + i \e 3 \f 3}{2}.
\end{equation}

The elements $\qqd$ and $\qdq$ are idempotent, since $(\qqd)^2=\qqd$ and $(\qdq)^2=\qdq$.
They are in fact primitive idempotent elements, hence they define minimal left and right ideals of the algebra $\SMA$ \citep{chevalley1997algebraicspinors,crumeyrolle1990clifford}.
When we represent the Clifford algebra $\SMA$ as an endomorphism algebra $\End_\C(\C^8)$, the idempotents $\qqd$ and $\qdq$ are represented as projectors.

The ideals $\SMA\qdvol=\SMA\qqd$ and $\SMA\qvol=\SMA\qdq$ are minimal left ideals, and the ideals $\qvol\SMA=(\SMA\qdvol)^\dagger$ and $\qdvol\SMA=(\SMA\qvol)^\dagger$ are minimal right ideals.
It is easy to show that $\ExtAlg\CSElectroColor\qvol=0$ and $\ExtAlg\CSElectroColor^\dagger\qdvol=0$, and therefore $\SMA\qdvol=\ExtAlg\CSElectroColor\qdvol=\ExtAlg\CSElectroColor^\dagger\qqd$ and $\SMA\qvol=\ExtAlg\CSElectroColor^\dagger\qvol=\ExtAlg\CSElectroColor\qdq$. Similar relations hold for the minimal right ideals, $\qvol\ExtAlg\CSElectroColor=0$, $\qdvol\ExtAlg\CSElectroColor^\dagger=0$, $\qvol\SMA=\qvol\ExtAlg\CSElectroColor^\dagger=\qqd\ExtAlg\CSElectroColor$, and $\qdvol\SMA=\qdvol\ExtAlg\CSElectroColor=\qdq\ExtAlg\CSElectroColor^\dagger$.

Let us describe the representation of the algebra $\SMA$ on its ideal $\ExtAlg\CSElectroColor^\dagger\qqd$.
The Clifford product between $u^\dagger + v\in\CSElectroColor^\dagger \oplus \CSElectroColor$ and $\omega\qqd\in\ExtAlg\CSElectroColor^\dagger\qqd$ is
\begin{equation}
\label{eq:sma_ext_representation}
(u^\dagger + v)\omega\qqd = (u^\dagger\wedge\omega)\qqd + (i_{v} \omega)\qqd \in\ExtAlg\CSElectroColor^\dagger\qqd,
\end{equation}
where the \emph{interior product} $i_{v} \omega$ is defined for any $\omega\in\ExtPow k\CSElectroColor^\dagger$ by
\begin{equation}
\label{eq:interior_product}
\(i_{v} \omega\)(u_1,\ldots,u_k) = 
\begin{cases}
\omega(v,u_1,\ldots,u_{k-1}), &\tn{ for } k\in\{1,2,3\}, \tn{ and} \\
0 &\tn{ for } k =0.
\end{cases}
\end{equation}

Then, the vectors $\q j$ act like \emph{annihilation operators}, and $\qd j$ act as \emph{creation operators} on $\ExtAlg\CSElectroColor^\dagger\qqd$:
\begin{equation}
\label{eq:color_creation_annihilation}
\begin{cases}
\qd j(\omega\qqd) = (\qd j \wedge \omega)\qqd,\\
\q j(\omega\qqd) = (i_{\q j} \omega)\qqd,
\end{cases}
\end{equation}
which is consistent with the anticommutation relations \eqref{eq:sma_anticommutation_relations}.

Similarly to equation \eqref{eq:sma_ext_representation} one defines an irreducible representation on the minimal left ideal $\ExtAlg\CSElectroColor\qdq$ of $\SMA$.
Its elements are $\CCl_6$-spinors.

A basis of the ideal $\ExtAlg\CSElectroColor^\dagger\qqd$ can be obtained from the basis \eqref{eq:basis_CSElectroColorExt},
\begin{equation}
\label{eq:left_ideal_basis_idempotent}
(1\ \qqd, \qd{23}\ \qqd, \qd{31}\ \qqd, \qd{12}\ \qqd, \qd{321}\ \qqd, \qd{1}\ \qqd, \qd{2}\ \qqd, \qd{3}\ \qqd).
\end{equation}

The basis \eqref{eq:left_ideal_basis_idempotent} is written in terms of the idempotent element $\qqd$. It is equal to the basis
\begin{equation}
\label{eq:left_ideal_basis_nilpotent}
(\qvol\ \qdvol, -\q{1}\ \qdvol, -\q{2}\ \qdvol, -\q{3}\ \qdvol, 1\ \qdvol, \q{23}\ \qdvol, \q{31}\ \qdvol, \q{12}\ \qdvol)
\end{equation}
written in terms of the nilpotent $\qdvol$, which determines the same ideal as $\qqd$.

Let us find the matrix representation of $\q j$, $\qd j$, $\e j$, and $\f j$ in the basis \eqref{eq:left_ideal_basis_idempotent}.

Here and in other places  it will be convenient to use the \emph{Pauli matrices} $\sigma_1=
{\scriptscriptstyle
	\left(\begin{array}{rr}
		0 & 1 \\
		1 & 0 \\
	\end{array}\right)}$,
$\sigma_2=
{\scriptscriptstyle
	\left(\begin{array}{rr}
		0 & -i \\
		i & 0 \\
	\end{array}\right)}$,
$\sigma_3=
{\scriptscriptstyle
	\left(\begin{array}{rr}
		1 & 0 \\
		0 & -1 \\
	\end{array}\right)}$,
and the matrices
$\sigma_+=\frac12(\sigma_1+i\sigma_2)=
{\scriptscriptstyle
	\left(\begin{array}{rr}
		0 & 1 \\
		0 & 0 \\
	\end{array}\right)}$,
$\sigma_-=\frac12(\sigma_1-i\sigma_2)=
{\scriptscriptstyle
	\left(\begin{array}{rr}
		0 & 0 \\
		1 & 0 \\
	\end{array}\right)}$,
$\sigma_3^+=\frac12(1+\sigma_3)=
{\scriptscriptstyle
	\left(\begin{array}{rr}
		1 & 0 \\
		0 & 0 \\
	\end{array}\right)}=\sigma_+\sigma_-$, and
$\sigma_3^-=\frac12(1-\sigma_3)=
{\scriptscriptstyle
	\left(\begin{array}{rr}
		0 & 0 \\
		0 & 1 \\
	\end{array}\right)}
	=\sigma_-\sigma_+$.

We obtain, in the representation \eqref{eq:left_ideal_basis_idempotent} of $\SMA$ on its ideal $\ExtAlg\CSElectroColor^\dagger\qvol$,
\begin{equation}
\label{eq:matrix_form_qdj}
	\qd 1=
{\scriptscriptstyle
	\left(\begin{array}{cccc}
		0 &  0 &  0 &  0 \\
		0 &  0 &  0 &  -i\sigma_2 \\
		-i\sigma_2 &  0 &  0 &  0 \\
		0 &  0 &  0 &  0 \\
	\end{array}\right)},
	\qd 2=
{\scriptscriptstyle
	\left(\begin{array}{cccc}
		0 &  0 &  0 &  \sigma_3^- \\
		0 &  0 &  -\sigma_3^- &  0 \\
		0 &  -\sigma_3^+ &  0 &  0 \\
		\sigma_3^+ &  0 &  0 &  0 \\
	\end{array}\right)},
	\qd 3=
{\scriptscriptstyle
	\left(\begin{array}{cccc}
		0 &  0 &  0 &  -\sigma_- \\
		0 &  0 &  \sigma_+ &  0 \\
		0 &  -\sigma_+ &  0 &  0 \\
		\sigma_- &  0 &  0 &  0 \\
	\end{array}\right)}.
\end{equation}

\begin{equation}
\label{eq:matrix_form_qj}
	\q 1=
{\scriptscriptstyle
	\left(\begin{array}{cccc}
		0 &  0 &  i\sigma_2 &  0 \\
		0 &  0 &  0 &  0 \\
		0 &  0 &  0 &  0 \\
		0 &  i\sigma_2 &  0 &  0 \\
	\end{array}\right)},
	\q 2=
{\scriptscriptstyle
	\left(\begin{array}{cccc}
		0 &  0 &  0 &  \sigma_3^+ \\
		0 &  0 &  -\sigma_3^+ &  0 \\
		0 &  -\sigma_3^- &  0 &  0 \\
		\sigma_3^- &  0 &  0 &  0 \\
	\end{array}\right)},
	\q 3=
{\scriptscriptstyle
	\left(\begin{array}{cccc}
		0 &  0 &  0 &  \sigma_+ \\
		0 &  0 &  -\sigma_- &  0 \\
		0 &  \sigma_- &  0 &  0 \\
		-\sigma_+ &  0 &  0 &  0 \\
	\end{array}\right)}.
\end{equation}

Then,
\begin{equation}
\label{eq:matrix_form_qvol}
	\qdvol=
{\scriptscriptstyle
	\left(\begin{array}{cccc}
		0 &  0 &  0 &  0 \\
		0 &  0 &  0 &  0 \\
		\sigma_3^+ &  0 &  0 &  0 \\
		0 &  0 &  0 &  0 \\
	\end{array}\right)},
	\qvol=
{\scriptscriptstyle
	\left(\begin{array}{cccc}
		0 &  0 &  \sigma_3^+ &  0 \\
		0 &  0 &  0 &  0 \\
		0 &  0 &  0 &  0 \\
		0 &  0 &  0 &  0 \\
	\end{array}\right)},
	\qqd=
{\scriptscriptstyle
	\left(\begin{array}{cccc}
		\sigma_3^+ &  0 &  0 &  0 \\
		0 &  0 &  0 &  0 \\
		0 &  0 &  0 &  0 \\
		0 &  0 &  0 &  0 \\
	\end{array}\right)},
	\qdq=
{\scriptscriptstyle
	\left(\begin{array}{cccc}
		0 &  0 &  0 &  0 \\
		0 &  0 &  0 &  0 \\
		0 &  0 &  \sigma_3^+ &  0 \\
		0 &  0 &  0 &  0 \\
	\end{array}\right)}.
\end{equation}

Then, from equation \eqref{eq:basis_sma},
\begin{equation}
\label{eq:matrix_form_ej}
	\e 1=
{\scriptscriptstyle
	\left(\begin{array}{cccc}
		0 &  0 &  i\sigma_2 &  0 \\
		0 &  0 &  0 &  -i\sigma_2 \\
		-i\sigma_2 &  0 &  0 &  0 \\
		0 &  i\sigma_2 &  0 &  0 \\
	\end{array}\right)},
	\e 2=
{\scriptscriptstyle
	\left(\begin{array}{cccc}
		0 &  0 &  0 &  1_2 \\
		0 &  0 &  -1_2 &  0 \\
		0 &  -1_2 &  0 &  0 \\
		1_2 &  0 &  0 &  0 \\
	\end{array}\right)},
	\e 3=
{\scriptscriptstyle
	\left(\begin{array}{cccc}
		0 &  0 &  0 &  i\sigma_2 \\
		0 &  0 &  i\sigma_2 &  0 \\
		0 &  -i\sigma_2 &  0 &  0 \\
		-i\sigma_2 &  0 &  0 &  0 \\
	\end{array}\right)},
\end{equation}
and
\begin{equation}
\label{eq:matrix_form_fj}
	\f 1=
{\scriptscriptstyle
	\left(\begin{array}{cccc}
		0 &  0 &  \sigma_2 &  0 \\
		0 &  0 &  0 &  \sigma_2 \\
		\sigma_2 &  0 &  0 &  0 \\
		0 &  \sigma_2 &  0 &  0 \\
	\end{array}\right)},
	\f 2=
{\scriptscriptstyle
	\left(\begin{array}{cccc}
		0 &  0 &  0 &  -i\sigma_3 \\
		0 &  0 &  i\sigma_3 &  0 \\
		0 &  -i\sigma_3 &  0 &  0 \\
		i\sigma_3 &  0 &  0 &  0 \\
	\end{array}\right)},
	\f 3=
{\scriptscriptstyle
	\left(\begin{array}{cccc}
		0 &  0 &  0 &  -i\sigma_1 \\
		0 &  0 &  i\sigma_1 &  0 \\
		0 &  -i\sigma_1 &  0 &  0 \\
		i\sigma_1 &  0 &  0 &  0 \\
	\end{array}\right)}.
\end{equation}

The matrix representations of the elements $\evol$, $\fvol$, and $\evol\fvol$ is
\begin{equation}
\label{eq:matrix_form_vol}
	\evol=
	{\scriptscriptstyle\left(\begin{array}{rr}
		0_4 &  1_4 \\
		-1_4 &  0_4 \\
	\end{array}\right)},
	\fvol=
	i
	{\scriptscriptstyle\left(\begin{array}{rr}
		0_4 &  1_4 \\
		1_4 &  0_4 \\
	\end{array}\right)},
	\evol\fvol=
	i
	{\scriptscriptstyle\left(\begin{array}{rr}
		1_4 &  0_4 \\
		0_4 &  -1_4 \\
	\end{array}\right)}.
\end{equation}

By noticing that $\qdvol \SMA\qvol = \C\qdvol\qvol$, it can be checked that
\begin{equation}
\label{eq:left_ideal_hermitian_metric}
\h(a,b)\qdq = (a^\dagger \qvol)^\dagger b^\dagger \qvol = \qdvol a b^\dagger\qvol
\end{equation}
for any $a,b\in\ExtAlg\CSElectroColor$. The proof is given in Appendix \sref{s:hermitian}. 
This shows that, although the inner product \eqref{eq:sma_inner_prod} vanishes on $\CSElectroColor^\dagger$ and $\CSElectroColor$, where the Hermitian inner product $\h$ is positive definite, the inner product $\h$ becomes unified with the Hermitian inner product in the ideal $\ExtAlg\CSElectroColor^\dagger\qvol$.

It is helpful sometimes to use multiindices $K\subset\{1,2,3\}$. This allows us to write immediately a matrix representation of the algebra $\SMA$.
We can represent the spinors from $\ExtAlg\CSElectroColor^\dagger\qvol$ as vectors
\begin{equation}
\label{eq:basis_sma_spinors}
\Psi\qvol=\sum_{K\subset\{1,2,3\}}\Psi^{K}\qd{K} \qvol,
\end{equation}
where $\Psi^{K}\in\C$.
Similarly, their duals can be expressed in the following vector form
\begin{equation}
\label{eq:basis_sma_cospinors}
\qdvol\Psi^\dagger=\sum_{K\subset\{1,2,3\}}\Psi^\dagger_{K}\qdvol\q{K},
\end{equation}
where $\Psi^\dagger_{K}\in\C$.

Then, the Hermitian scalar product \eqref{eq:left_ideal_hermitian_metric}
\begin{equation}
\label{eq:hermitian_metric}
\h(\Phi,\Psi)\qdq=\(\Phi\qvol\)^\dagger\Psi\qvol
\end{equation}
takes the form
\begin{equation}
\label{eq:vector_hermitian_metric}
\h(\Phi,\Psi)=\sum_{K\subset\{1,2,3\}}\Phi^\dagger_{K}\Psi^{K}.
\end{equation}

Any element $a$ of $\SMA$ can be written uniquely as a linear combination of the form
\begin{equation}
\label{eq:basis_sma_matrix}
a=\sum_{K_1,K_2\subset\{1,2,3\}}a^{K_1}{}_{K_2}\ \qd{K_1} \ \qqd \ \q{K_2}.
\end{equation}
Then, from equation \eqref{eq:vector_hermitian_metric} follows that for any $a,b\in\SMA$, the Clifford product $c=ab$ takes the simple matrix form
\begin{equation}
\label{eq:basis_sma_matrix_multiplication}
c^{K_1}{}_{K_2}=\sum_{K\subset\{1,2,3\}}a^{K_1}{}_{K}b^{K}{}_{K_2},
\end{equation}
for all $K_1,K_2\subset\{1,2,3\}$.

Therefore, the Witt decomposition $\SMA^1=\CSElectroColor^\dagger \oplus \CSElectroColor$ gives a natural decomposition of $\SMA$ as a direct sum of left ideals
\begin{equation}
\label{eq:sma_ideal_decomposition}
\SMA=\bigoplus_{k=0}^{3} \(\ExtAlg\CSElectroColor^\dagger\)\qqd\ExtPow k\CSElectroColor,
\end{equation}
which means that $\SMA$ decomposes as sum of spinors with internal degrees of freedom in $\ExtPow k\CSElectroColor$, similar to leptons and quarks, as explained in Section \sref{s:patternsSM}.

%-----------------------------------------------------%
\section{The weak symmetry}
\label{s:weak_symmetry}

In this section I review the weak symmetry, and after that I show how it is present in the algebra $\SMA$.

%-----------------------------------------------------%
\subsection{Review of the weak symmetry}
\label{s:weak_symmetry_review}

For generality, I will consider a pair of $4$-spinors $\WeakUp\in\wsu$ and $\WeakDown\in\wsd$, belonging to distinct representations of the Dirac algebra $\wsu$ and $\wsd$, such that that their left-handed components $\WeakUpL\in\wsuL$ and $\WeakDownL\in\wsdL$ belong to the same weak doublet. In particular, the pair
${\scriptscriptstyle
	\left(\begin{array}{l}
		\WeakUp \\
		\WeakDown \\
	\end{array}\right)}$
can be leptons like
${\scriptscriptstyle
	\left(\begin{array}{l}
		\nu_e \\
		e^- \\
	\end{array}\right)}$,
	quarks of the same color like
${\scriptscriptstyle
	\left(\begin{array}{l}
		u_r \\
		d_r \\
	\end{array}\right)}$,
and so on for the other families. At this stage we will focus on the $\SymWeak$ symmetry, and ignore other degrees of freedom, which will emerge naturally later.
Also, for simplicity here I discuss only the weak symmetry and not the electroweak symmetry, which is the true gauge symmetry. The electroweak symmetry will be discussed later.

The Dirac matrices act simultaneously and identically on the $4$-spinors $\WeakUp$ and $\WeakDown$. 
The chiral components of the two $4$-spinors are
\begin{equation}
\label{eq:chiral_components}
\begin{cases}
\WeakUpL := P_L \WeakUp\in\wsuL, & \WeakUpR = P_R \WeakUp\in\wsuR, \\
\WeakDownL = P_L \WeakDown\in\wsdL, & \WeakDownR = P_R \WeakDown\in\wsdR, \\
\end{cases}
\end{equation}
where
\begin{equation}
\label{eq:chiral_projectors}
\begin{cases}
P_L  := \frac{1 - \gamma^5}{2}, \\
P_R  := \frac{1 + \gamma^5}{2}. \\
\end{cases}
\end{equation}
We denote the chiral subspaces of the complex four-dimensional spaces $\wsu$ and $\wsd$ by
\begin{equation}
\label{eq:chiral_subspaces}
\begin{cases}
\wsuL := P_L \wsu, & \wsuR = P_R \wsu, \\
\wsdL = P_L \wsd, & \wsdR = P_R \wsd. \\
\end{cases}
\end{equation}

Weak interactions distinguish between left handed and right handed leptons and quarks. 
The generators of the weak symmetry group $\SymWeak$ are defined on the left chiral components by
\begin{equation}
\label{eq:SymWeak-generators_L}
\begin{array}{l}
T_{jL}:\End_\C(\wsuL\oplus\wsdL), \\
T_{jL} := \sigma_j\otimes 1_2, \\
\end{array}
\end{equation}
where $j\in\{1,2,3\}$ and $\sigma_j$ are the Pauli matrices.
In other words,
\begin{equation}
\begin{cases}
T_{1L}
{\scriptscriptstyle
\left(\begin{array}{r}
		\WeakUpL \\
		\WeakDownL \\
	\end{array}\right)}
=
{\scriptscriptstyle
\left(\begin{array}{r}
		\WeakDownL \\
		\WeakUpL \\
	\end{array}\right)}, \\
T_{2L}
{\scriptscriptstyle
\left(\begin{array}{r}
		\WeakUpL \\
		\WeakDownL \\
	\end{array}\right)}
=
{\scriptscriptstyle
\left(\begin{array}{r}
		-i\WeakDownL \\
		i\WeakUpL \\
	\end{array}\right)}, \\
T_{3L}
{\scriptscriptstyle
\left(\begin{array}{r}
		\WeakUpL \\
		\WeakDownL \\
	\end{array}\right)}
=
{\scriptscriptstyle
\left(\begin{array}{r}
		\WeakUpL \\
		-\WeakDownL \\
	\end{array}\right)}.
\end{cases}
\end{equation}

On the right chiral components they act by
\begin{equation}
\label{eq:SymWeak-generators_R}
\begin{array}{l}
T_{jR}:\End_\C(\wsuR\oplus\wsdR), \\
T_{jR} := 0 \\
\end{array}
\end{equation}
for all $j\in\{1,2,3\}$.
They generate the trivial action of $\SymWeak$ on the right-handed fermions, so
\begin{equation}
T_{jR}{\scriptscriptstyle
\left(\begin{array}{r}
		\WeakUpR \\
		\WeakDownR \\
	\end{array}\right)}=0.
\end{equation}

It follows that the generators of the weak symmetry on the total spinor space $\wsu\oplus\wsd$ are $T_j = T_{jL}\oplus T_{jR}$. More explicitly,
\begin{equation}
\label{eq:SymWeak-generators}
\begin{array}{l}
T_j:\End_\C(\wsu\oplus\wsd), \\
T_j = \sigma_j \otimes P_L, \\
\end{array}
\end{equation}
where $j\in\{1,2,3\}$.

Hence, 
\begin{equation}
\label{eq:SymWeak-generators_commutators}
\frac12[T_j,T_k] = i\epsilon^{jkl}T_l,
\end{equation}
where $j,k,l\in\{1,2,3\}$. 
The structure constants $\epsilon^{jkl}$ are completely antisymmetric, being equal to $0$ except for $\{j,k,l\}=\{1,2,3\}$, in which case $\epsilon^{jkl}=\pm1$, the parity of the permutation of the indices $j,k,l$.

In Section \sref{s:from_weak_to_ccl_six} I explain how the Dirac algebra combined with the weak symmetry generate the algebra $\Matrix8\C\cong\CCl_6$.

%-----------------------------------------------------%
\subsection{Weak symmetry in the algebra \texorpdfstring{$\SMA$}{A}}
\label{s:weak_symmetry_sma}

Consider now again the minimal left ideal $\SMA\qqd$ of $\SMA$. It is an eight-dimensional complex vector space. By right multiplication with various exterior products of the elements $\q j$, one gets ideals of charges that are multiple of $\frac13$ and colors. But to represent leptons and quarks, one has to find out how the Dirac algebra acts on each of these ideals. Clearly the representation of the Dirac algebra on an eight-dimensional complex vector space is reducible, but the projectors $\frac12\(1\pm i\evol\fvol\)$ provide the decomposition into irreducible representations. In addition, each of the resulting four-dimensional subspaces has to be split into complex two-dimensional spaces corresponding to chirality. So we need the representation of the chirality operator, which we take to be
\begin{equation}
\label{eq:chirality_operator}
\tgamma^5 := -i\e1\f1=
{\scriptscriptstyle
	\left(\begin{array}{cccc}
	 1_2 &  0 &  0 &  0 \\
		0 &  -1_2 &  0 &  0 \\
		0 &  0 &  -1_2 &  0 \\
		0 &  0 &  0 & 1_2 \\
	\end{array}\right)}.
\end{equation}
This choice favors a particular color -- direction in the space of colors. However, we will see later that this choice does not break the $\SymColor$ symmetry, because the action of the Dirac algebra on the ideals is independent of the colors and the action of $\SymColor$.

Now we look for the representations of $\SymWeak$, taking into account the chirality of each space.
To find it, consider the elements
\begin{equation}
\label{eq:matrix_form_bwdj}
	\bw 1=
{\scriptscriptstyle
	\left(\begin{array}{cccc}
		0 &  1_2 &  0 &  0 \\
		0 &  0 &  0 &  0 \\
		0 &  0 &  0 &  -1_2 \\
		0 &  0 &  0 &  0 \\
	\end{array}\right)},
	\bw 2=
{\scriptscriptstyle
	\left(\begin{array}{cccc}
		0 &  0 &  -1_2 &  0 \\
		0 &  0 &  0 &  -1_2 \\
		0 &  0 &  0 &  0 \\
		0 &  0 &  0 &  0 \\
	\end{array}\right)},
	\bw 3=
{\scriptscriptstyle
	\left(\begin{array}{cccc}
		\sigma_+  &  0 &  0 &  0 \\
		0 &  -\sigma_+  &  0 &  0 \\
		0 &  0 &  -\sigma_+  &  0 \\
		0 &  0 &  0 &  \sigma_+  \\
	\end{array}\right)}.
\end{equation}

The reader may anticipate that the indices $u$ and $d$ are related to the weak symmetry doublets. Indeed, the reasons for this notation will become apparent, and also the necessity of $\bw3$ and $\bwd3$, which are not related to the weak symmetry.

We define the null complex vector spaces $\ns$ and $\nsd$ as the spaces spanned by null vectors from \eqref{eq:basis_w}, by
\begin{equation}
\label{eq:w_spaces}
\begin{cases}
\ns := \Span_\C\(\bw1,\bw2,\bw3\), \\
\nsd := \Span_\C\(\bwd1,\bwd2,\bwd3\). \\
\end{cases}
\end{equation}

The elements
\begin{equation}
\label{eq:basis_w}
(\bw1, \bw2, \bw3, \bwd1, \bwd2, \bwd3)
\end{equation}
form a \emph{Witt basis} of the space $\nsd\oplus\ns$, satisfying the \emph{anticommutation relations}
\begin{equation}
\label{eq:w_anticommutation_relations}
\begin{array}{l}
\{\bw j, \bw k\} = 0, \\
\{\bwd j, \bwd k\} = 0, \\
\{\bw j, \bwd k\} = \delta_{jk} \\
\end{array}
\end{equation}
for $j,k\in\{\bwindex1,\bwindex2,\bwindex3\}$.

We define the orthonormal basis
\begin{equation}
\label{eq:basis_w_forward}
\begin{cases}
\we j = \bw j + \bwd j \\
\wf j = i\(\bwd j - \bw j\), \\
\end{cases}
\end{equation}
where $j\in\{\bwindex1,\bwindex2,\bwindex3\}$.
Then, $\we j^2=1$, $\wf j^2=1$, and
\begin{equation}
\label{eq:basis_w_back}
\begin{cases}
\bw j = \frac12\(\we j + i\wf j\) \\
\bwd j = \frac12\(\we j - i\wf j\). \\
\end{cases}
\end{equation}

The matrix form of $\we j$ and $\wf j$ is
\begin{equation}
\label{eq:matrix_form_wej}
	\we 1=
{\scriptscriptstyle
	\left(\begin{array}{cccc}
		0 &  1_2 &  0 &  0 \\
		1_2 &  0 &  0 &  0 \\
		0 &  0 &  0 &  -1_2 \\
		0 &  0 &  -1_2 &  0 \\
	\end{array}\right)},
	\we 2=
{\scriptscriptstyle
	\left(\begin{array}{cccc}
		0 &  0 &  -1_2 &  0 \\
		0 &  0 &  0 &  -1_2 \\
		-1_2 &  0 &  0 &  0 \\
		0 &  -1_2 &  0 &  0 \\
	\end{array}\right)},
	\we 3=
{\scriptscriptstyle
	\left(\begin{array}{cccc}
		\sigma_1  &  0 &  0 &  0 \\
		0 &  -\sigma_1  &  0 &  0 \\
		0 &  0 &  -\sigma_1  &  0 \\
		0 &  0 &  0 &  \sigma_1  \\
	\end{array}\right)},
\end{equation}
\begin{equation}
\label{eq:matrix_form_wfj}
	\wf 1=
{\scriptscriptstyle
	\left(\begin{array}{cccc}
		0 &  -i1_2 &  0 &  0 \\
		i1_2 &  0 &  0 &  0 \\
		0 &  0 &  0 &  i1_2 \\
		0 &  0 &  -i1_2 &  0 \\
	\end{array}\right)},
	\wf 2=
{\scriptscriptstyle
	\left(\begin{array}{cccc}
		0 &  0 &  i1_2 &  0 \\
		0 &  0 &  0 &  i1_2 \\
		-i1_2 &  0 &  0 &  0 \\
		0 &  -i1_2 &  0 &  0 \\
	\end{array}\right)},
	\wf 3=
{\scriptscriptstyle
	\left(\begin{array}{cccc}
		\sigma_2  &  0 &  0 &  0 \\
		0 &  -\sigma_2  &  0 &  0 \\
		0 &  0 &  -\sigma_2  &  0 \\
		0 &  0 &  0 &  \sigma_2  \\
	\end{array}\right)}.
\end{equation}

It is useful to notice that $\we 1 =-i\e3\f1$, $\wf 1 =\fvol\e2$, $\we 2 =i\fvol$, and $\wf 2 =i\evol$.
\begin{equation}
\label{eq:wewg-vs-ef}
	\left\{\begin{array}{llll}
		\we 1 &=-i\e3\f1, &  \wf 1 &=\fvol\e2, \\
		\we 2 &=i\fvol, &  \wf 2 &=i\evol. \\
	\end{array}\right.
\end{equation}

It is easy to check that none of the elements $\we j,\wf j,\bwd j,\bw j$ are linear combinations of the elements $(\qd 1,\qd 2,\qd 3,\q 1,\q 2,\q 3)$.
It follows that the complex vector space $\nsd\oplus\ns$ is different from $\CSElectroColor^\dagger \oplus \CSElectroColor$.

The elements 
\begin{equation}
\label{eq:bwvol}
\begin{cases}
\bwvol :=& \bw 1 \bw 2 \bw 3,\\
\bwdvol =& \bwd 3 \bwd 2 \bwd 1 \\
\end{cases}
\end{equation}
are \emph{nilpotent}, since $\bwvol^2=0$ and $\bwdvol{}^2=0$.

The nilpotents $\bwvol$ and $\bwdvol$ have the following matrix form
\begin{equation}
\label{eq:matrix_form_nilpotents}
	\bwvol=
{\scriptscriptstyle
	\left(\begin{array}{cccc}
		0 &  0 &  0 &  -\sigma_+ \\
		0 &  0 &  0 &  0 \\
		0 &  0 &  0 &  0 \\
		0 &  0 &  0 &  0 \\
	\end{array}\right)},
	\bwdvol=
{\scriptscriptstyle
	\left(\begin{array}{cccc}
		0 &  0 &  0 &  0 \\
		0 &  0 &  0 &  0 \\
		0 &  0 &  0 &  0 \\
		-\sigma_- &  0 &  0 &  0 \\
	\end{array}\right)}.
\end{equation}

From them we can construct the \emph{idempotents} $\bwdvol\bwvol$ and $\bwvol\bwdvol$,
\begin{equation}
\label{eq:matrix_form_idempotents}
	\bwvol\bwdvol=\wproj=
{\scriptscriptstyle
	\left(\begin{array}{cccc}
		\sigma_3^+ &  0 &  0 &  0 \\
		0 &  0 &  0 &  0 \\
		0 &  0 &  0 &  0 \\
		0 &  0 &  0 &  0 \\
	\end{array}\right)},
	\bwdvol\bwvol=
{\scriptscriptstyle
	\left(\begin{array}{cccc}
		0 &  0 &  0 &  0 \\
		0 &  0 &  0 &  0 \\
		0 &  0 &  0 &  0 \\
		0 &  0 &  0 &  \sigma_3^- \\
	\end{array}\right)}.
\end{equation}

Then,
\begin{equation}
\label{eq:weak_ideal}
\SMA\wproj = \ExtAlg \nsd\wproj.
\end{equation}

By identifying the complex three-dimensional vector space $\nsd$ with the dual of $\ns$, the anticommutation relations \eqref{eq:w_anticommutation_relations} are equivalent to a Hermitian inner product $\hns$ defined similarly to \eqref{eq:left_ideal_hermitian_metric}.

The vectors $\bw j$ and $\bwd j$ act as ladder operators on this ideal, similar to \eqref{eq:color_creation_annihilation}:
\begin{equation}
\label{eq:w_creation_annihilation}
\begin{cases}
\bwd j(a\wproj) = (\bwd j \wedge a)\wproj,\\
\bw j(a\wproj) = (i_{\bw j} a)\wproj,
\end{cases}
\end{equation}
where $a\in\ExtAlg \nsd$, and $i_{\bw j}$ is the interior product defined by the Hermitian inner product $\hns$.
This definition is consistent with the anticommutation relations \eqref{eq:w_anticommutation_relations}.

From the relations \eqref{eq:w_creation_annihilation} it follows that the matrix form \eqref{eq:matrix_form_bwdj} corresponds to the basis
\begin{equation}
\label{eq:basis_weak}
\(1\ \wproj,\bwd3 \ \wproj,\bwd1 \ \wproj,\bwd1 \bwd3 \ \wproj,\bwd2 \ \wproj,\bwd2 \bwd3 \ \wproj, \bwd2 \bwd1 \ \wproj,\bwd2 \bwd1 \bwd3 \ \wproj\).
\end{equation}

At the same time, the matrices \eqref{eq:matrix_form_bwdj} are expressed in the basis \eqref{eq:left_ideal_basis_idempotent}. Hence,
\begin{equation}
\label{eq:basis_weak_color_relation}
\left\{\begin{array}{rrr}
\bwd3 \ \wproj &= &\qd{23}\ \qqd \\
\bwd1 \ \wproj &= &\qd{31}\ \qqd \\
\bwd1 \bwd3 \ \wproj &= &\qd{12}\ \qqd \\
\bwd2 \ \wproj &= &\qd{321}\ \qqd \\
\bwd2 \bwd3 \ \wproj &= &\qd{1}\ \qqd \\
\bwd2 \bwd1 \ \wproj &= &\qd{2}\ \qqd \\
\bwd2 \bwd1 \bwd3 \ \wproj &= &\qd{3}\ \qqd \\
\end{array}\right.
\end{equation}
We cannot conclude, for example from $\bwd3 \wproj = \qd{23}\qqd$, that $\bwd3 = \qd{23}$, because we are not allowed to divide by $\wproj$, which is not invertible. Although the identities \eqref{eq:basis_weak_color_relation} are between elements of the same ideal $\ExtAlg \nsd\wproj = \ExtAlg\CSElectroColor^\dagger\qqd$, the spaces $\ExtAlg \nsd$ and $\ExtAlg\CSElectroColor^\dagger$ are different.

%-----------------------------------------------------%
\subsection{Spinorial generators of the weak symmetry}
\label{s:weak_symmetry_generators}

The key to the weak symmetry is the two-dimensional Hermitian vector space 
\begin{equation}
\label{eq:weak_space}
\(\wsw:=\Span_\C\(\bwd1, \bwd2\), \hw\),
\end{equation}
where $\hw:=\hns|_{\wsw}$ is the restriction of $\hns$ to $\wsw$.

The exterior powers $\ExtPow 0\wsw = \Span_\C\(1\)$ and $\ExtPow 2\wsw = \Span_\C\(\bwd1 \bwd2\)$ represent the singlet states, and $\ExtPow 1\wsw = \Span_\C\(\bwd1, \bwd2\)$ the doublet.

But in order to represent the elementary fermions of the Standard Model, the singlet spaces and the doublet space need to have twice the number of dimensions of the exterior powers $\ExtPow k\wsw$. This is obtained by the additional $\bwd 3$, which also is needed to represent the Dirac algebra. In addition, we need more degrees of freedom to represent not only  the leptons, but also the up and down quarks, and their colors. These will emerge naturally as other ideals of $\SMA$.

Let $\wsuR:=\Span_\C\(\wproj,\bwd3 \wproj\)$ be the vector subspace of the ideal $\SMA\wproj$ spanned by the null vectors $\wproj$ and $\bwd3 \wproj$.
In the following, it will correspond to the \emph{up} particle singlet space of the weak symmetry.
The elements of the basis \eqref{eq:basis_weak} split the ideal $\SMA\wproj$ into subspaces which correspond to the singlets and doublets of the weak symmetry:
\begin{equation}
\label{eq:weak_spaces}
\begin{cases}
\tn{Right-handed up singlet space: } \wsuR := 1\Span_\C\(\wproj,\bwd3 \wproj\), \\
\tn{Left-handed up doublet space: } \wsuL := \bwd1 \wsuR, \\
\tn{Right-handed down singlet space: } \wsdR := \bwd1 \bwd2  \wsuR, \\
\tn{Left-handed down doublet space: } \wsdL := \bwd2  \wsuR. \\
\end{cases}
\end{equation}

The Clifford algebra $\SMA$ contains a spin representation of the weak group $\SymWeak$, which is a double cover of the representation normally used.

We choose the following set of generator bivectors for the group $\SymWeak$:
\begin{equation}
\label{eq:weak_generators}
\begin{cases}
\wg1 := \we 1 \wf 2 - \wf 1 \we 2 \\
\wg2 := \we 1 \we 2 + \wf 1 \wf 2 \\
\wg3 := \we 1 \wf 1 - \we 2 \wf 2 \\
\end{cases}
\end{equation}

They have the following matrix form in the basis \eqref{eq:basis_weak}:
\begin{equation}
\label{eq:matrix_form_wgj}
	\wg1=
2i{\scriptscriptstyle
	\left(\begin{array}{cccc}
		0 &  0 &  0 &  0 \\
		0 &  0 &  1_2 &  0 \\
		0 &  1_2 &  0 &  0 \\
		0 &  0 &  0 &  0 \\
	\end{array}\right)},
	\wg2=
2{\scriptscriptstyle
	\left(\begin{array}{cccc}
		0 &  0 &  0 &  0 \\
		0 &  0 &  -1_2 &  0 \\
		0 &  1_2 &  0 &  0 \\
		0 &  0 &  0 &  0 \\
	\end{array}\right)},
	\wg3=
2i{\scriptscriptstyle
	\left(\begin{array}{cccc}
		0  &  0 &  0 &  0 \\
		0 &  -1_2  &  0 &  0 \\
		0 &  0 &  1_2  &  0 \\
		0 &  0 &  0 &  0 \\
	\end{array}\right)}.
\end{equation}

Let us check that the bivectors in equation \eqref{eq:weak_generators} are spinorial generators of the $\SymWeak$ group.
We notice that each generator is a sum of commuting pure bivectors, $\wg k = B_{1 k} + B_{2 k}$.
All these bivectors have negative square, $B_{j k}^2 < 0$. Because of commutativity of the bivectors in each generator, the generated group elements are of the form
\begin{equation}
e^{\frac12\varphi\wg k} = e^{\frac12\varphi B_{1 k}} e^{\frac12\varphi B_{2 k}},
\end{equation}
where the factors commute, and $\varphi\in\R$.

Let us see how the transformations generated by $\wg 1$, $\wg 2$, and $\wg 3$ act on $\bwd 1$, $\bwd 2$, and $\bwd 3$. From \eqref{eq:weak_generators} we see that all of the generators commute with $\we3$ and $\wf3$, hence also with $\bwd 3$.

The transformations generated by $\wg 1$ act on $\bwd j$ like
\begin{equation}
\label{eq:weak_generators-a}
\begin{array}{ll}
e^{\frac12\varphi\wg 1} \bwd j e^{-\frac12\varphi\wg 1}
&= \frac12 e^{\frac12\varphi\wg 1} (\we j - i\wf j) e^{-\frac12\varphi\wg 1} \\
&= \frac12 e^{\frac12\varphi \we 1 \wf 2} e^{-\frac12\varphi \wf 1 \we 2} (\we j - i\wf j) e^{\frac12\varphi \wf 1 \we 2}e^{-\frac12\varphi \we 1 \wf 2}. \\
\end{array}
\end{equation}
Then, for $\bwd 1$, $\bwd 2$, and $\bwd 3$,
\begin{equation}
\begin{array}{ll}
e^{\frac12\varphi\wg 1} \bwd 1 e^{-\frac12\varphi\wg 1}
&= \frac12 \(e^{\frac12\varphi \we 1 \wf 2} \we 1 e^{-\frac12\varphi \we 1 \wf 2}
- i e^{-\frac12\varphi \wf 1 \we 2} \wf 1 e^{\frac12\varphi \wf 1 \we 2}\) \\
&= \frac12 \(e^{\varphi \we 1 \wf 2} \we 1
- i e^{-\varphi \wf 1 \we 2} \wf 1 \) \\
&= \frac12 \( \we 1\cos \varphi - \wf 2\sin \varphi \) - \frac12 i \( \wf 1\cos \varphi + \we 2\sin \varphi \) \\
&= \frac12 \( \we 1 - i \wf 1\) \cos \varphi - \frac12 i \(\we 2 - i \wf 2\)\sin \varphi \\
&= \bwd 1\cos \varphi - i \bwd 2\sin \varphi. \\
\end{array}
\end{equation}

\begin{equation}
\begin{array}{ll}
e^{\frac12\varphi\wg 1} \bwd 2 e^{-\frac12\varphi\wg 1}

&= \frac12 \(e^{-\frac12\varphi \wf 1 \we 2} \we 2 e^{\frac12\varphi \wf 1 \we 2}
- i e^{\frac12\varphi \we 1 \wf 2} \wf 2 e^{-\frac12\varphi \we 1 \wf 2}\) \\
&= \frac12 \( e^{-\varphi \wf 1 \we 2} \we 2
- i e^{\varphi \we 1 \wf 2} \wf 2 \) \\
&= \frac12 \( \we 2\cos \varphi - \wf 1\sin \varphi \) - \frac12 i \( \wf 2\cos \varphi + \we 1\sin \varphi \) \\
&= \frac12 \( \we 2 - i \wf 2\) \cos \varphi - \frac12 i \(\we 1 - i \wf 1\)\sin \varphi \\
&= \bwd 2\cos \varphi - i \bwd 1\sin \varphi. \\
\end{array}
\end{equation}

\begin{equation}
\begin{array}{ll}
e^{\frac12\varphi\wg 1} \bwd 3 e^{-\frac12\varphi\wg 1}
&= \bwd 3. \\
\end{array}
\end{equation}
Therefore, for any $a$ in the plane $\wsw$ from \eqref{eq:weak_space},
\begin{equation}
\label{eq:weak_generators-up-down-plane-a}
e^{\frac\varphi 2 \wg 1} a e^{-\frac\varphi 2 \wg 1} = e^{-i\varphi\sigma_1} a.
\end{equation}

The transformations generated by $\wg 2$ act on $\bwd j$ like
\begin{equation}
\label{eq:weak_generators-b}
\begin{array}{ll}
e^{\frac12\varphi\wg 2} \bwd j e^{-\frac12\varphi\wg 2}
&= \frac12 e^{\frac12\varphi\wg 2} (\we j - i\wf j) e^{-\frac12\varphi\wg 2} \\
&= \frac12 \(e^{\frac12\varphi \we 1 \we 2} \we j e^{-\frac12\varphi \we 1 \we 2} - i e^{\frac12\varphi \wf 1 \wf 2} \wf j e^{-\frac12\varphi \wf 1 \wf 2}\) \\
&= 
\begin{cases}
\frac12 e^{\varphi \we 1 \we 2} \we j - \frac12 i e^{\varphi \wf 1 \wf 2} \wf j &\tn{ if } j\in\{\bwindex1,\bwindex2\}\\
\we j &\tn{ if } j=\bwindex3.\\
\end{cases}
\end{array}
\end{equation}
Then, for $\bwd 1$, $\bwd 2$, and $\bwd 3$,
\begin{equation}
\begin{array}{ll}
e^{\frac12\varphi\wg 2} \bwd 1 e^{-\frac12\varphi\wg 2}
&= \frac12 \(e^{\varphi \we 1 \we 2} \we 1 - i e^{\varphi \wf 1 \wf 2}\wf 1\) \\
&= \frac12 \( \we 1\cos \varphi - \we 2\sin \varphi \) - \frac12 i \( \wf 1\cos \varphi - \wf 2\sin \varphi \) \\
&= \bwd 1\cos \varphi - \bwd 2\sin \varphi. \\
\end{array}
\end{equation}

\begin{equation}
\begin{array}{ll}
e^{\frac12\varphi\wg 2} \bwd 2 e^{-\frac12\varphi\wg 2}
&= \frac12 \(e^{\varphi \we 1 \we 2} \we 2 - i e^{\varphi \wf 1 \wf 2}\wf 2\) \\
&= \frac12 \( \we 2\cos \varphi + \we 1\sin \varphi \) - \frac12 i \( \wf 2\cos \varphi + \wf 1\sin \varphi \) \\

&= \bwd 1\sin \varphi + \bwd 2\cos \varphi. \\
\end{array}
\end{equation}

\begin{equation}
\begin{array}{ll}
e^{\frac12\varphi\wg 2} \bwd 3 e^{-\frac12\varphi\wg 2}
&= \bwd 3. \\
\end{array}
\end{equation}
Hence, for any $a$ in the plane $\wsw$,
\begin{equation}
\label{eq:weak_generators-up-down-plane-b}
e^{\frac\varphi 2 \wg 2} a e^{-\frac\varphi 2 \wg 2} = e^{-i\varphi\sigma_2} a.
\end{equation}

The transformations generated by $\wg 3$ are
\begin{equation}
\label{eq:weak_generators-c}
\begin{array}{ll}
e^{\frac12\varphi\wg 3} \bwd j e^{-\frac12\varphi\wg 3}
&= e^{\frac12\varphi \we 1 \wf 1} e^{-\frac12\varphi \we 2 \wf 2} \bwd j e^{\frac12\varphi \we 2 \wf 2}e^{-\frac12\varphi \we 1 \wf 1}. \\
\end{array}
\end{equation}
Therefore,
\begin{equation}
\begin{array}{ll}
e^{\frac12\varphi\wg 3} \bwd 1 e^{-\frac12\varphi\wg 3} = e^{\varphi \we 1 \wf 1} \bwd 1,\\
e^{\frac12\varphi\wg 3} \bwd 2 e^{-\frac12\varphi\wg 3} = e^{-\varphi \we 2 \wf 2} \bwd 2,\\
e^{\frac12\varphi\wg 3} \bwd 3 e^{-\frac12\varphi\wg 3} = \bwd 3.\\
\end{array}
\end{equation}
Again, for any $a$ in the plane $\wsw$,
\begin{equation}
\label{eq:weak_generators-up-down-plane-c}
e^{\frac\varphi 2 \wg 3} a e^{-\frac\varphi 2 \wg 3} = e^{-i\varphi\sigma_3} a.
\end{equation}

Using \eqref{eq:weak_spaces}, we summarize the action of $\wg j$ from \eqref{eq:weak_generators-up-down-plane-a}, \eqref{eq:weak_generators-up-down-plane-b}, \eqref{eq:weak_generators-up-down-plane-c}, and the results that these transformations leave $\bwd3$ and $\bwd1\bwd2$ unchanged, in the following:
\begin{equation}
\label{eq:weak_generators-up-down-plane}
\begin{cases}
e^{\frac\varphi 2 \wg j} a e^{-\frac\varphi 2 \wg j} = e^{-i\varphi\sigma_j} a,&\tn{ for any $a$ in $\Span_\C(\bwd1,\bwd2)$ and $\Span_\C(\bwd1\bwd3,\bwd2\bwd3)$},\\
e^{\frac\varphi 2 \wg j} a e^{-\frac\varphi 2 \wg j} = a,&\tn{ for any $a$ in $\Span_\C(1,\bwd3,\bwd1\bwd2,\bwd1\bwd2\bwd3)$},\\
\end{cases}
\end{equation}
where $j\in\{\bwindex1,\bwindex2,\bwindex3\}$.

By \eqref{eq:weak_spaces}, the action from \eqref{eq:weak_generators-up-down-plane} becomes
\begin{equation}
\label{eq:weak_generators-spin-two}
\begin{cases}
e^{\frac\varphi 2 \wg j} a e^{-\frac\varphi 2 \wg j} = e^{-i\varphi\sigma_j\otimes1_2} a,&\tn{ for any $a\in\wsuL\oplus\wsdL$},\\
e^{\frac\varphi 2 \wg j} a e^{-\frac\varphi 2 \wg j} = a,&\tn{ for any $a\in\wsuR\oplus\wsdR$},\\
\end{cases}
\end{equation}
or, if we combine them,
\begin{equation}
\label{eq:weak_generators-spin}
e^{\frac\varphi 2 \wg j} a e^{-\frac\varphi 2 \wg j} = e^{-i\varphi\sigma_j\otimes P_L} a
\end{equation}
for any $a\in\wsuL\oplus\wsdL$ and $j\in\{1,2,3\}$.

The action of the operators \eqref{eq:weak_generators} on the basis \eqref{eq:basis_weak} of the space $\ExtAlg \nsd=\wsu\oplus\wsd$ shows that they generate an $\SU(2)$ symmetry. 
The relation with the generators of the $\SymWeak$ symmetry from \eqref{eq:SymWeak-generators} is
\begin{equation}
\label{eq:weak_generators-spin-SymWeak}
e^{-i\varphi T_j} a = e^{\frac\varphi 2 \wg j} a e^{-\frac\varphi 2 \wg j},
\end{equation}
for any $a\in\wsuL\oplus\wsdL$ and $j\in\{1,2,3\}$.

Hence, we recovered the usual generators of the $\SymWeak$ group from the spin group generators \eqref{eq:weak_generators}.

%-----------------------------------------------------%
\section{The Dirac algebra}
\label{s:dirac_algebra}

While we can use any other basis, I preferred the one from \eqref{eq:basis_weak}, because it is easily related to the Weyl basis for the Dirac matrices, and to the usual way to describe the electroweak interaction. 

Let us recall the chiral (Weyl) representation,
\begin{equation}
\label{eq:dirac_matrix_weyl}
	\gamma^0=
	{\scriptscriptstyle\left(\begin{array}{cc}
		0 &  1_2 \\
		1_2 &  0 \\
	\end{array}\right)},
	\gamma^j=
	{\scriptscriptstyle\left(\begin{array}{cc}
		0 &  \sigma_j \\
		-\sigma_j &  0 \\
	\end{array}\right)},
	\gamma^5=
	{\scriptscriptstyle\left(\begin{array}{cc}
		-1_2 &  0 \\
		0 &  1_2 \\
	\end{array}\right)}
\end{equation}
and define a modified version of it
\begin{equation}
\label{eq:dirac_matrix_weyl_mod}
	\gammamod^0=
	{\scriptscriptstyle\left(\begin{array}{cc}
		0 &  1_2 \\
		1_2 &  0 \\
	\end{array}\right)},
	\gammamod^j=
	{\scriptscriptstyle\left(\begin{array}{cc}
		0 &  -\sigma_j \\
		\sigma_j &  0 \\
	\end{array}\right)},
	\gammamod^5=
	{\scriptscriptstyle\left(\begin{array}{cc}
		1_2 &  0 \\
		0 &  -1_2 \\
	\end{array}\right)}.
\end{equation}

As we can see from the spinorial representation of the $\SymWeak$ group, the Dirac representation on the eight-dimensional space $\SMA\wproj$ is the direct sum of the two chiral representations,
\begin{equation}
\label{eq:dirac_matrix_t}
	\tgamma^{\mu}=
	{\scriptscriptstyle\left(\begin{array}{cc}
		\gammamod^{\mu} &  0 \\
		0 &  \gamma^{\mu} \\
	\end{array}\right)}.
\end{equation}

From \eqref{eq:matrix_form_wej} and \eqref{eq:matrix_form_wfj} follows that, in the basis \eqref{eq:basis_weak},
\begin{equation}
\label{eq:matrix_form_splitters}
	i\we 1\wf 1=
{\scriptscriptstyle
	\left(\begin{array}{cccc}
		-1_2 &  0 &  0 &  0 \\
		0 &  1_2 &  0 &  0 \\
		0 &  0 &  -1_2 &  0 \\
		0 &  0 &  0 &  1_2 \\
	\end{array}\right)},
	i\we 2\wf 2=
{\scriptscriptstyle
	\left(\begin{array}{cccc}
		-1_2 &  0 &  0 &  0 \\
		0 &  -1_2 &  0 &  0 \\
		0 &  0 &  1_2 &  0 \\
		0 &  0 &  0 &  1_2 \\
	\end{array}\right)},
	i\we 3\wf 3=
-{\scriptscriptstyle
	\left(\begin{array}{cccc}
		\sigma_3 &  0 &  0 &  0 \\
		0 &  \sigma_3 &  0 &  0 \\
		0 &  0 &  \sigma_3 &  0 \\
		0 &  0 &  0 &  \sigma_3 \\
	\end{array}\right)}.
\end{equation}

It follows that
\begin{equation}
\label{eq:tgamma_five}
\tgamma^5 =
{\scriptscriptstyle\left(\begin{array}{cc}
		\gammamod^5 &  0 \\
		0 & \gamma^5 \\
	\end{array}\right)}
	=\we 1\we 2\wf 1\wf 2.
\end{equation}

%In the chiral (Weyl) basis LR, the charge conjugation operator is $\mc C\psi = -i\gamma^2\psi^\ast$.

%-----------------------------------------------------%
\section{The electroweak symmetry}
\label{s:electroweak}

In the standard electroweak theory, the electromagnetic and weak interactions are an artifact of the broken electroweak symmetry, and are considered less fundamental than the electroweak interaction. In the approach proposed here the electromagnetic and weak interactions seem to be more fundamental than the electroweak interaction, and that the hypercharge is less fundamental than the electric charge and the weak isospin.
I do not exclude the possibility of a reconstruction of the Standard Model from the algebra $\SMA$ starting with the electroweak symmetry first, followed by a symmetry breaking into the electromagnetic and weak symmetries. But we will see that the electroweak symmetry breaking appears to come from the geometry of the algebra $\SMA$, rather than being spontaneous.
The electroweak symmetry is still present in the approach proposed here, but the electromagnetic and weak symmetries are distinguished by the geometry.

In the following I discuss the electroweak symmetry breaking from geometric point of view. I will review first the geometry of the standard electroweak symmetry breaking in a way similar to (\cite{Derdzinski1992GeometryStandardModel}, Ch. 6). Then, I will calculate the Weinberg angle as seems to be predicted by the algebra $\SMA$.

The exchange bosons of the electroweak force are connections in the gauge bundle having as fiber the two-dimensional Hermitian vector space $\(\wsw, \hw\)$, where $\wsw:=\Span_\C\(\bwd1, \bwd2\)$ (equation \eqref{eq:weak_space}). Consequently, the internal components of the exchange bosons of the electroweak force are elements of the unitary Lie algebra $\ssymElectroWeak\cong\uu\(\wsw\)$, that is, Hermitian forms. The unitary Lie algebra $\ssymElectroWeak$, regarded as a vector space, has four real dimensions. After the symmetry breaking, they correspond to the photon $\gamma$, and the weak force bosons $W^\pm$ and $Z^0$. Following \citep{Derdzinski1992GeometryStandardModel}, the decomposition of the Lie algebra $\ssymElectroWeak$ into subspaces where each of these bosons live is
\begin{equation}
\label{eq:ew_bosons_spaces}
\uu\(\wsw\) = \gamma\(\wsw\) \oplus W\(\wsw\) \oplus Z\(\wsw\).
\end{equation}
Hence, $\gamma\in\gamma\(\wsw\)$, $W^\pm\in W\(\wsw\)$, and $Z^0\in Z\(\wsw\)$.
The decomposition \eqref{eq:ew_bosons_spaces} is not unique, but is uniquely determined by the \emph{Higgs field} $\phi$ and the \emph{Weinberg electroweak mixing angle} $\theta_W$.
In fact, what we need is a special complex line in the space $\wsw$, which is determined by $\phi$, and an Ad-invariant inner product on $\uu\(\wsw\)$.
The requirement that the inner product is invariant results in the following form:
\begin{equation} 
\label{eq:hermitian-forms-inner-product-ew}
\langle a,b\rangle_{\uu\(\wsw\)} = -2r_2g'{}^2 \Tr(ab) + r_2(g'{}^2-g^2)\Tr a\Tr b,
\end{equation} 
where $a,b\in\uu\(\wsw\)$, $g,g'$ are constants -- the coupling constants of the electroweak model, and $r_2>0$ is a constant.
The Weinberg angle $\theta_W$ is given by
\begin{equation} 
\label{eq:weinberg-angle-def}
\sin^2\theta_W=\frac{g'{}^2}{g^2+g'{}^2}.
\end{equation} 

The electric charge $e$ is
\begin{equation} 
\label{eq:electric-charge}
e = g\sin\theta_W=g'\cos\theta_W=\frac12\sqrt{g^2+g'{}^2}\sin2\theta_W.
\end{equation} 

The standard electroweak model does not provide a preference for this angle, which is determined indirectly from experiments.
The grand unified theories, and the present proposal, predict definite values for the Weinberg angle.

The Higgs field is a scalar with respect to spacetime symmetries, but internally it is a vector $\phi\in\wsw$. The direction of the vector $\phi$ in $\wsw$ is the element $\bwd1=\frac{\phi}{\sqrt{\hw(\phi,\phi)}}$.
The Higgs field has two main roles: on the one hand is responsible for the symmetry breaking, by selecting a particular direction in the space $\wsw$.
On the other hand, it is responsible for the masses of at least some of the elementary particles.

The Higgs field is a section of the electroweak bundle, which splits the electroweak bundle for a pair of weakly interacting leptons into two one-dimensional complex bundles -- the bundle spanned by the Higgs field, and the bundle orthogonal to that. But in the proposed approach, this split is ensured by the operator $-i\evol\fvol$. Recall that the representation of the Dirac algebra on one of the minimal left ideals of $\SMA$ is reducible, being eight-dimensional. The operator $-i\evol\fvol$ splits each ideal into two four-dimensional space by determining two projectors, $\frac12\(1\mp i\evol\fvol\)$. Therefore, it also determines the particular direction $\bwd1$, and by this, the Higgs field $\phi$ up to a constant factor. Hence, in the algebra $\SMA$, the symmetry breaking does not require the Higgs field, although it is still needed to generate the masses of the particles.

Let us now calculate the prediction of the Weinberg angle $\theta_W$, first in general, considering an extension of $\uu(2)$ to $\su(N)$, $2<N\in\N$. I will follow a simple generalization of the usual geometric proof, used for example in (\cite{Derdzinski1992GeometryStandardModel}, Ch. 7) for the $\SU(5)$ GUT.
Because $\su(N)$ is simple, there is a unique Ad-invariant inner product, up to a constant $r$,
\begin{equation} 
\label{eq:hermitian-forms-inner-product-suN}
\langle A,B\rangle_{\SU(N)} = - N r_N \Tr(AB),
\end{equation} 
where $A,B\in\su(N)$, $r_N>0$. The embedding of $\uu(2)$ in $\su(N)$ should be traceless, because $\Tr(A)=0$ for any $A\in\su(N)$.
It follows that the embedding is given, in a basis extending the basis of $\wsw$ to $\C^N$. by
\begin{equation} 
\label{eq:hermitian-forms-extension-ew-to-suN}
a\mapsto a \oplus \(-\frac1{N-2}\Tr a I_{\wsw^\perp}\)
\end{equation}
for any $a\in\uu(2)$.

Then,
\begin{equation} 
\label{eq:hermitian-forms-inner-product--ew-to-suN-calculation}
\begin{array}{ll}
\langle a,b\rangle_{\uu(2)}
&=\langle a \oplus \(-\frac1{N-2}\Tr a I_{\wsw^\perp}\),b \oplus \(-\frac1{N-2}\Tr b I_{\wsw^\perp}\)\rangle_{\SU(N)}\\
&= - N r_N \Tr(ab) - N r_N \(-\frac1{N-2}\)^2\Tr a \Tr b \Tr I_{\wsw^\perp} \\
&= - N r_N \Tr(ab) - N r_N \frac{1}{N-2}\Tr a \Tr b . \\ 
\end{array}
\end{equation}
By comparing with \eqref{eq:hermitian-forms-inner-product-ew} it follows that $2r_2g'{}^2=r_NN$ and $r_2(g'{}^2-g^2) = -N r_N\frac{1}{N-2}$.
This solves to $g'{}^2 = \frac{N}{2}\frac{r_N}{r_2}$ and $g^2 = g'{}^2 + \frac{N}{N-2}\frac{r_N}{r_2}$.
Then, the Weinberg angle predicted by a GUT based on the extension of $\uu(2)$ to $\su(N)$ is
\begin{equation} 
\label{eq:weinberg-angle-N}
\sin^2\theta_{W,N}=\frac{\frac{N}{2}}{N + \frac{N}{N-2}} = \frac{\frac{N}{2}}{\frac{N(N-1)}{N-2}} = \frac12 \frac{N-2}{N-1}.
\end{equation} 

Then, for the $\SU(5)$ GUT model one gets $\sin^2\theta_{W,5} = \frac38 = 0.375$.

For the algebra $\SMA$, recall that $\uu\(\wsw\)$ is embedded in $\su(3)$, which is the symmetry group of $(\nsd,\hns)$. Then,
\begin{equation} 
\label{eq:weinberg-angle-3}
\sin^2\theta_{W,\SMA}=\sin^2\theta_{W,3}=\frac14 = 0.25,
\end{equation} 
corresponding to $\theta_{W,\SMA} = \frac\pi6$.

The prediction of $\SMA$, $\sin^2\theta_W=0.25$, seems more encouraging that that of $0.375$ predicted by the $\SU(5)$, $\Spin(10)$, and other GUTs.
But its derivation from the embedding of $\SymElectroWeak$ into an $\SU(3)$ symmetry acting on the left of the algebra $\SMA$ seems to imply an unexpected connection between the electroweak symmetry and spacetime, which requires further investigations.
Moreover, it is still not within the range estimated experimentally.
Depending on the utilized scheme, the experimental values for $\sin^2\theta_W$, range between $\sim 0.223$ and $\sim 0.24$ \citep{PDG2015ElectroWeakConstraints}. In particular, CODATA gives a value of $0.231 29(5)$ \citep{PDG2016PhysicalConstants}. As in the case of the $\SU(5)$ prediction of $\sin^2\theta_{W,5}=0.375$, a correct comparison would require taking into account the running of the coupling constants due to higher order perturbative corrections.
Note that there are other models which predict the same value for $\theta_W$, in particular in \citep{Besprosvany2000GaugeSpaceTimeUnification,Daviau2015RetourALOndeDeLouisDeBroglie,Daviau2015WeinbergAngle}.

A Weinberg angle fixed by the algebraic structure used in a particular model may seem to be a problem, for two reasons. On the one hand the value has to somehow fit within the experimental bounds, which may be explained by higher order corrections. On the other hand, the model has to explain the anomaly cancellation, which is perfectly well explained within the electroweak symmetry breaking. A fixed geometric or algebraic structure, which gives a fixed value for this angle, can still be compatible with renormalization, because there are other parameters that come into the equations. For example, there are many reasons to believe that renormalization in both quantum field theory and quantum gravity is related to \emph{dimensional reduction} effects of one kind or another. A possible reason for this dependance of the energy of the coupling constants and related parameters as proposed in several models was suggested in \citep{Sto12d,Sto13a} to be due to general relativistic effects accompanying spacetime singularities. Whether this may solve the problems related to a fixed Weinberg angle in this model is open for future investigations.

%-----------------------------------------------------%
\section{The electrocolor symmetry}
\label{s:gauge_electrocolor}

We will see now that the algebra $\SMA$, which has the symmetries of the gauge groups $\SymColor$ and $\SymEM$, contains the generators of these groups. The groups $\SymColor$ and $\SymEM$ are subgroups of $\SO(\CSElectroColor^\dagger \oplus \CSElectroColor)$, which is double-covered by the group $\Spin(\CSElectroColor^\dagger \oplus \CSElectroColor)$,

The \emph{complex special orthogonal Lie algebra} $\so(\CSElectroColor^\dagger \oplus \CSElectroColor)\cong\so(6,\C) \cong \spin(\CSElectroColor^\dagger \oplus \CSElectroColor) \cong \spin(6,\C)$ corresponding to the inner product from equation \eqref{eq:sma_inner_prod} is the Lie algebra $\(\SMA^2,[\cdot,\cdot]\)$, where $[a,b] := ab-ba$ for any $a,b\in\SMA^2$, so it consists of the bivectors of $\SMA$. Since the subspace $\CSElectroColor$ is isotropic with respect to the inner product \eqref{eq:sma_inner_prod}, the \emph{complex general linear group} $\GL(\CSElectroColor)$ is a subgroup of $\SO(\CSElectroColor^\dagger \oplus \CSElectroColor)$. The special unitary group $\SymColor = \SU(\CSElectroColor,\h)$ is in turn a subgroup of $\GL(\CSElectroColor)$, so its Lie algebra $\su(\CSElectroColor,\h)$ is a Lie subalgebra of $\(\SMA^2,[\cdot,\cdot]\)$.

A set of generator bivectors for the group $\SymColor$ can be chosen to correspond to the \emph{Gell-Mann matrices},
\begin{equation}
\label{eq:color_generators}
\begin{array}{lll}
\tlambda_1 = \e 1 \f 2 - \f 1 \e 2, &\tlambda_2 = \e 1 \e 2 + \f 1 \f 2, &\tlambda_3 = \e 1 \f 1 - \e 2 \f 2, \\
\tlambda_4 = \e 1 \f 3 - \f 1 \e 3, &\tlambda_5 = \e 1 \e 3 + \f 1 \f 3, & \\
\tlambda_6 = \e 2 \f 3 - \f 2 \e 3, &\tlambda_7 = \e 2 \e 3 + \f 2 \f 3, &\tlambda_8 = \frac{1}{\sqrt{3}} (\e 1 \f 1 + \e 2 \f 2 - 2 \e 3 \f 3). \\
\end{array}
\end{equation}

The proof is similar to that for the $\SymWeak$ symmetry given in Section \sref{s:weak_symmetry_generators}.

It is immediate to see that $\tlambda_j^\dagger=-\tlambda_j$ for all values of $j$.

The standard Gell-Mann matrices are defined by $\lambda_j=i\tlambda_j$. Then,
\begin{equation}
\label{eq:color_generators-spin-SymColor}
e^{-i\varphi \lambda_j} a = e^{\frac\varphi 2 \tlambda j} a e^{-\frac\varphi 2 \tlambda j},
\end{equation}
for the $\SymColor$ representation $\mathbf{3}$.

As in the case of the $\SymWeak$ symmetry, what we did was to obtain the double cover of $\SymColor$, which turns out to be a subgroup of the spin group $\Spin(\CSElectroColor^\dagger \oplus \CSElectroColor)$. The action of an element $g\in\Spin(\CSElectroColor^\dagger \oplus \CSElectroColor)$ on an element $\omega\in\SMA$ is given by $\omega \mapsto g\omega g^{-1}$.
This means that the action of $\SymColor$ on $\CSElectroColor$ extends to the exterior algebra $\ExtAlg\CSElectroColor$, in a way compatible with the exterior product.
It follows that the spinorial action generated by the elements $\tlambda_j$ is equivalent to the right multiplication with the matrix $O_1\oplus e^{i\varphi \lambda_j}\oplus O_1\oplus e^{-i\varphi \lambda_j}$. This corresponds to the representations $\mathbf{1}_c$, $\mathbf{3}_c$, $\overline{\mathbf{1}}_c$, and $\overline{\mathbf{3}}_c$.
Consequently, the action of $\SymColor$ on $\ExtAlg\CSElectroColor$ and on $\ExtAlg\CSElectroColor^\dagger$ is the one prescribed in Section \sref{s:patternsSM}, as it should.

Since the space $\CSElectroColor$ not only includes the color, but also the electric charge, let us find the generator of the electromagnetic group $\SymEM$.
Since the $\SymEM$ gauge transformation only multiplies the vectors in $\CSElectroColor$ by a phase factor $e^{i\varphi}$, it follows that the generator is the identity of $\End_\C\(\CSElectroColor\)$,
\begin{equation}
\label{eq:em_generator}
Q = \e 1 \f 1 + \e 2 \f 2 + \e 3 \f 3.
\end{equation}

Again, since we are using a spin representation, the action of the group $\SymEM$ on $\ExtAlg\CSElectroColor$ and on $\ExtAlg\CSElectroColor^\dagger$ is consistent with the exterior product, and the electric charge is proportional with the degree, so it is $\frac{k}{3}\mathit{e}$, where $\mathit{e}$ is the electron charge, and $k\in\{\pm0,\pm1,\pm2,\pm3\}$, with the identification $\ExtPow{-k}\CSElectroColor=\ExtPow k\CSElectroColor^\dagger=\ExtPow k\CSAElectroColor$.

Equations \eqref{eq:color_generators} and \eqref{eq:em_generator} demonstrate how the color and electromagnetic symmetries are unified into an \emph{electrocolor} symmetry $\U(3)_{ec}$.

The symmetry generated by \eqref{eq:em_generator} transforms not only $\qqd\ExtPow k\CSElectroColor$, but also $\bwd 2\wproj$. From $\bwd 2\wproj =\qdvol\wproj$ it follows that the electric charge of $\qdvol\wproj$ is $-1$.
This accounts for the fact that each minimal left ideal contains two different particles, with different electric charges.

I arrived at the symmetries $\SymColor$ and $\SymEM$ and the generators \eqref{eq:color_generators} and \eqref{eq:em_generator} starting from the standard ideal decomposition of Clifford algebras $\CCl_{2r}$ \citep{chevalley1997algebraicspinors,crumeyrolle1990clifford}, the representation of $\U(N)$ and $\SU(N)$ on $\Cl_{2N}$ as the subgroup of $\Spin(2N)$ preserving a Hermitian inner product, given in \citep{DoranHestenes1993LGasSG}, and by the standard construction of the Hermitian exterior algebra \citep{ROWells2007ComplexManifolds}, resulting in the correct $\mathbf{1}_c,\mathbf{3}_c,\overline{\mathbf{1}}_c$, ad $\overline{\mathbf{3}}_c$ representations.
A proof that the unitary spin transformations preserving a Witt decomposition in $\CCl_6$ give the $\SymColor$ and $\SymEM$ symmetries, along with a set of generators constructed from the $\q j$ and $\qd j$ ladder operators but equivalent to \eqref{eq:color_generators}, was given in \citep{Furey2015ChargeQuantization}.
Based on the algebra $\Cl_7$, in \citep{TraylingBaylis2004Cl7StandardModel} were proposed generators of $\SymColor$ which are equivalent to \eqref{eq:color_generators} due to the isomorphisms $\Cl_7\cong\Matrix8\C\cong\CCl_6$.

%-----------------------------------------------------%
\section{Leptons and quarks}
\label{s:gauge_leptons_quarks}

We have seen in Section \sref{s:sma_ideal} that the algebra $\SMA$ decomposes naturally into eight minimal left ideals \eqref{eq:sma_ideal_decomposition},
\begin{equation}
\label{eq:sma_ideal_color}
\SMA=\bigoplus_{k=0}^{3} \SMA\qqd\ExtPow k\CSElectroColor.
\end{equation}

In Section \sref{s:weak_symmetry_generators} we have seen that the ideal $\SMA\wproj$ admits a decomposition compatible with the weak symmetry,
\begin{equation}
\label{eq:sma_ideal_weak}
\SMA\wproj = \wsu\oplus\wsd = \wsuR\oplus\wsuL\oplus\wsdL\oplus\wsdR.
\end{equation}

From \eqref{eq:sma_ideal_color} and \eqref{eq:sma_ideal_weak}, it follows that the $\SMA$ decomposes naturally into representations of the Dirac algebra
\begin{equation}
\label{eq:sma_ideal_weak_color_dirac}
\SMA=\bigoplus_{k=0}^{3} \(\wsu\oplus\wsd\)\ExtPow k\CSElectroColor,
\end{equation}
and into chiral spaces,
\begin{equation}
\label{eq:sma_ideal_weak_color_chiral}
\SMA=\bigoplus_{k=0}^{3} \(\wsuR\oplus\wsuL\oplus\wsdL\oplus\wsdR\)\ExtPow k\CSElectroColor.
\end{equation}

We proceed to identify each representation of the Dirac algebra from the decomposition \eqref{eq:sma_ideal_weak_color_dirac} of the algebra $\SMA$ with leptons and quarks. This identification should be made taking into account the types of charges and of gauge symmetries for each lepton and quark in a generic family.

First, we need to separate the internal and the external degrees of freedom which identify each of the subspaces from \eqref{eq:sma_ideal_weak_color_chiral}. We identify as \emph{external degrees of freedom} those that change under a Lorentz transformation, and as \emph{internal degrees of freedom} those that do not change.
It follows that the internal degrees of freedom, those that classify the decomposition \eqref{eq:sma_ideal_weak_color_chiral}, are labeled by $\bwd 1$ and $\bwd 2$ -- corresponding to the weak symmetry, and by the various combinations of $\q K$, $K\subseteq\{1,2,3\}$ -- corresponding to the color. In fact, $\bwd 1$ is not purely internal, since it changes under improper Lorentz transformations like time inversion.

The Dirac representations for leptons and quarks have the form $\wsu\ExtPow k\CSElectroColor$ and $\wsd\ExtPow k\CSElectroColor$, or, in other words, $\wsu\q K$ and $\wsd\q K$, where $K\subseteq\{1,2,3\}$.
If the set $K$ has one element, one gets the elements $\q1,\q2,\q3$, which have electric charge $+\frac13$ and colors $\overline{\mathbf{r}},\overline{\mathbf{y}}$, and $\overline{\mathbf{b}}$. If $K$ has two elements, one obtains $\q{23},\q{31},\q{12}$ have electric charge $+\frac23$ and colors $\mathbf{r},\mathbf{y}$, and $\mathbf{b}$.
Taking into account that $\bwd 2\wproj=\we 2\wproj$ and that $\we2=i\fvol$,
\begin{equation}
\bwd 2\wproj =\we 2\wproj= i\fvol\wproj=\qdvol\wproj.
\end{equation}
Then, if $\ws{}=\Span_\C\(1,\bwd3,\bwd1,\bwd1\bwd3\)$, one obtains
\begin{equation}
\begin{cases}
\wsu=\ws{} \wproj,\\
\wsd=\ws{}\bwd 2\wproj=\ws{}\we2\wproj=\ws{}\qdvol\wproj,
\end{cases}
\end{equation}

The electric charge of $\qdvol=\(\q1\q2\q3\)^\dagger$ is $-1$, and it has no color.
The element $\wproj=\qvol\qdvol$ is electrically neutral and without color.
The product $\qdvol\wproj$ has therefore the electric charge $-1$ and no color.
The elements $\bwd 1\wproj$ and $\bwd 2\wproj$ are invariant to proper Lorentz transformations, and due to the spinorial representation of $\SymWeak$ from Section \sref{s:weak_symmetry_generators}, they have \emph{weak isospin $I_3$} of $+\frac12$ and $-\frac12$. The element $\bwd 1\bwd 2\wproj$ has the weak isospin equal to $0$.

Therefore, the ideal $\(\wsu\oplus\wsd\)\ExtPow 0\CSElectroColor$ represents the neutrino and the electron, while $\(\wsu\oplus\wsd\)\ExtPow 2\CSElectroColor$ represents up and down quarks.
What about the ideals $\(\wsu\oplus\wsd\)\ExtPow 1\CSElectroColor$ and $\(\wsu\oplus\wsd\)\ExtPow 3\CSElectroColor$? They seem to correspond, by their charges, to particles similar to the down and up antiquarks, and respectively to the positron and antineutrino. But they should not be new particles, they should be antiparticles of the leptons and quarks.
To understand this, we take a closer look at the structure of an ideal $\(\wsu\oplus\wsd\)\ExtPow k\CSElectroColor$. Then,
\begin{equation}
\label{eq:ideals-particle-antiparticle}
\begin{array}{ll}
\(\wsu\oplus\wsd\)\ExtPow k\CSElectroColor &= \(\ExtAlg \nsd\)\wproj\ExtPow k\CSElectroColor \\
&= \(\ExtAlg \nsd\)\qvol\qdvol\ExtPow k\CSElectroColor \\
&=\(\ExtAlg \nsd\)\qvol\qdvol\qvol\qdvol\ExtPow k\CSElectroColor \\
&= \(\ExtAlg \ns\)\qdvol\qvol\ExtPow {3-k}\CSAElectroColor\\
&= \(\ExtAlg \ns\)\qdq\ExtPow {3-k}\CSAElectroColor\\
&= \(\overline{\wsd}\oplus\overline{\wsu}\)\qdq\ExtPow {3-k}\CSAElectroColor.
\end{array}
\end{equation}
Hence, the ideal $\(\wsu\oplus\wsd\)\ExtPow k\CSElectroColor$ represents the antiparticles of the particles represented by the ideal $\(\wsu\oplus\wsd\)\ExtPow {3-k}\CSElectroColor$.
The action of the groups $\SymEM$, $\SymColor$, and $\SymWeak$, on the antiparticle ideals are opposite to those on the corresponding particle ideals.
The chiral spaces are reversed, and so is the basis of the weak charges.
For antiparticles the weak interaction takes place only between the components of right chirality. 

The \emph{hypercharge} $Y$ is obtained from the electric charge and the weak isospin by the \emph{Gell-Mann--Nishijima formula}
\begin{equation}
\label{eq:Gell-Mann-Nishijima}
Y = 2(Q-T_3).
\end{equation}

From \eqref{eq:weak_generators} and \eqref{eq:wewg-vs-ef} follows that 
\begin{equation}
\label{eq:weak_generators_b}
\begin{cases}
\wg1 = -\f1\e1\e2 - i\e2,\\
\wg2 = \e3\f2\f3 + i\fvol\e3\e1,\\
\wg3 = i\e2\e3\f2\f3 + \fvol\evol.\\
\end{cases}
\end{equation}
Hence, the spinorial generators of the electroweak symmetry \eqref{eq:weak_generators_b} do not commute with those of the color symmetry \eqref{eq:color_generators}.
However, the electroweak symmetry, like the Dirac algebra and the Lorentz group, acts on the ideals, while the color symmetry permutes the ideals.
Hence, their actions are independent. This is illustrated in \eqref{eq:standard_model_algebra} by the fact that some of them act on the rows, while the others on the columns, therefore commuting.

We centralize all these remarks, and use as classifiers the elements of the form $\wproj\q{K}$ and $\qdvol\wproj\q{K}$.
Then, the data in Table \ref{tab:standard_model_fermions} can be classified as in Table \ref{tab:standard_model_fermions_sma}.

\def\arraystretch{1.25}%  1 is the default, change whatever you need
\begin{table}[ht]
\begin{center}
    \begin{tabular}{ ? l l ? r | r | r | r ? p{3cm} |}
    \bottomrule[1.5pt]
		\rowcolor{gray!50}
    \textbf{Particle} & & $\mathbf{\nu}$ & $\mathbf{\overline d}$ & $\mathbf{u}$ & $\mathbf{e^+}$ \\\hline
    \textbf{Spinor space} & & $\wsu$ & $\wsu\q{j}$ & $\wsu\q{jk}$ & $\wsu\q{123}$ \\ \hline
    \textbf{Classifier} & & $\wproj$ & $\wproj\q{j}$ & $\wproj\q{jk}$ & $\wproj\q{123}$ \\ \hline
    \textbf{Electric charge} & & $0$ & $+\frac{1}{3}$ & $+\frac{2}{3}$ & $+1$ \\ \hline
		\multirow{2}{*}{\textbf{Chiral space}}
		& \textbf{L} & $\bwd1\wsu$ & $\wsu\q{j}$ & $\bwd1\wsu\q{jk}$ & $\wsu\q{123}$ \\
		& \textbf{R} & $\wsu$ & $\bwd1\wsu\q{j}$ & $\wsu\q{jk}$ & $\bwd1\wsu\q{123}$ \\ \hline
		\multirow{2}{*}{\textbf{Weak isospin}}
		& \textbf{L} & $+\frac{1}{2}$ & $0$ & $+\frac{1}{2}$ & $0$ \\
		& \textbf{R} & $0$ & $+\frac{1}{2}$ & $0$ & $+\frac{1}{2}$ \\ \hline
		\multirow{2}{*}{\textbf{Hypercharge}}
		& \textbf{L} & $-1$ & $+\frac{2}{3}$ & $+\frac{1}{3}$ & $+2$ \\
		& \textbf{R} & $0$ & $-\frac{1}{3}$ & $+\frac{4}{3}$ & $+1$ \\ \hline
		\rowcolor{gray!50}
    \textbf{Particle} & & $\mathbf{e^-}$ & $\mathbf{\overline u}$ & $\mathbf{d}$ & $\mathbf{\overline \nu}$ \\\hline
    \textbf{Spinor space} & & $\wsd$ & $\wsd\q{j}$ & $\wsd\q{jk}$ & $\wsd\q{123}$ \\ \hline
    \textbf{Classifier} & & $\qdvol\wproj$ & $\qdvol\wproj\q{j}$ & $\qdvol\wproj\q{jk}$ & $\qdvol\wproj\q{123}$ \\ \hline
    \textbf{Electric charge} & & $-1$ & $-\frac{2}{3}$ & $-\frac{1}{3}$ & $0$ \\ \hline
		\multirow{2}{*}{\textbf{Chiral space}}
		& \textbf{L} & $\bwd2\wsu$ & $\bwd1\bwd2\wsu\q{j}$ & $\bwd2\wsu\q{jk}$ & $\bwd1\bwd2\wsu\q{123}$ \\
		& \textbf{R} & $\bwd1\bwd2\wsu$ & $\bwd2\wsu\q{j}$ & $\bwd1\bwd2\wsu\q{jk}$ & $\bwd2\wsu\q{123}$ \\ \hline
		\multirow{2}{*}{\textbf{Weak isospin}}
		& \textbf{L} & $-\frac{1}{2}$ & $0$ & $-\frac{1}{2}$ & $0$ \\
		& \textbf{R} & $0$ & $-\frac{1}{2}$ & $0$ & $-\frac{1}{2}$ \\ \hline
		\multirow{2}{*}{\textbf{Hypercharge}}
		& \textbf{L} & $-1$ & $-\frac{4}{3}$ & $+\frac{1}{3}$ & $0$ \\
		& \textbf{R} & $-2$ & $-\frac{1}{3}$ & $-\frac{2}{3}$ & $1$ \\ \hline
		\toprule[1.25pt]
    \end{tabular}
\end{center}
\caption{Discrete properties of leptons and quarks in the algebra $\SMA$.}
\label{tab:standard_model_fermions_sma}
\end{table}

The leptons and quarks, as well as their antiparticles, sit therefore in the algebra $\SMA$ as in \eqref{eq:standard_model_algebra}.

%-----------------------------------------------------%
\section{All symmetries}
\label{s:sma-symmetries}

The algebra $\SMA$ includes the leptons and quarks from a generic family, as well as the electromagnetic, color, weak, and Lorentz symmetries.
In this section we will look at its symmetries. In order to do this, we have to review the layers of structures that define this algebra.

The first structure is isomorphic to the algebra $\Matrix8\C$, or the algebra of complex linear endomorphisms of an eight-dimensional complex vector space.
The next level structure is a special space of operators -- the subspace $\CSElectroColor^\dagger \oplus \CSElectroColor$ of $\Matrix8\C$.
This determines a gradation on $\Matrix8\C$, which makes it into the Clifford algebra $\CCl_8$.
On top of the Clifford algebra structure lies the Witt decomposition $\CSElectroColor^\dagger \oplus \CSElectroColor$.
From this, one obtains the electric and color charges, as well as the $\SymEM$ and $\SymColor$ symmetries, in the form of representations of subgroups of the spin group $\Spin(\CSElectroColor^\dagger \oplus \CSElectroColor)$.
These symmetries are the internal symmetries of the decomposition of $\CCl_8$ into eight complex eight-dimensional minimal left ideals.
Each of these ideals is characterized by an electric charge and color charge, which may be white.
The action of the Dirac algebra on each of these ideals gives a reducible representation. The representation is decomposed into irreducible representations by the projectors $\frac12\(1\mp i\evol\fvol\)$ determined by the volume element $\evol\fvol$ of $\CCl_8$.
On the reducible eight-dimensional representation of the Dirac algebra, that is, on each of these ideals, the weak symmetry generators act at left, also as generators of a subgroup of the spin group, but this time associated to a different subspace $\nsd\oplus\ns$ of $\CCl_8$. They gives the usual left action of the weak force generators. The Dirac matrices also act at left, and so do the Lorentz group.
The generators of the electromagnetic symmetry act both at right, like those of the color symmetry, and at left.

To see the way these various actions are nested on the representation of the algebra $\SMA$, let us recall that $\wsd=\ws{}\bwd 2\wproj=\ws{}\we2\wproj=\ws{}\qdvol\wproj=\ws{}\evol\wproj$. By this and \eqref{eq:sma_ideal_weak_color_chiral}, any element of the $\SMA$ is a linear combination of elements of the form
\begin{equation}
\label{eq:generic-element-nested-symmetries}
\bwd3{}^{a}\,\bwd1{}^b\,\qdvol{}^c\,\wproj\,\q{K},
\end{equation}
where $K\subset\{1,2,3\}$ is a multiindex, $a,b,c\in\{0,1\}$, and by convention, $\(\bwd1\)^0=\(\bwd3\)^0=\(\qdvol\)^0=1$.
Table \ref{tab:standard_model_sym} contains the ranges of action of each of the symmetry groups and the Dirac algebra. As we have seen, the weak symmetry group, the Dirac algebra and the Lorentz group act at the left of the projector operator $\wproj$, the color symmetry group at the right, and the electromagentic symmetry group on the right and partially on the left, but its action does not overlap with the external symmetries.

\def\arraystretch{1.5}%  1 is the default, change whatever you need
\begin{table}[ht]
\begin{center}
\begin{tabular}{?c|c|c?c?c?}
\bottomrule[1.5pt]\rowcolor{gray!50}
\hline
$\bwd3{}^{a}$ & $\bwd1{}^b$ & $\qdvol{}^c$ & $\wproj$ & $\q{K}$ \\ \hline
\multicolumn{2}{?c}{$\underbrace{\hspace{100pt}}_{\tn{Dirac algebra}}$} & \multicolumn{3}{c?}{$\underbrace{\hspace{158pt}}_{\SymEM}$} \\
\multicolumn{2}{?c}{$\underbrace{\hspace{100pt}}_{\GO(1,3)}$} &\multicolumn{1}{c}{}& \multicolumn{2}{c?}{$\underbrace{\hspace{100pt}}_{\SymColor}$} \\
\multicolumn{1}{?c}{$\underbrace{\hspace{45pt}}_{\SO^+(1,3)}$} & \multicolumn{3}{c}{$\underbrace{\hspace{158pt}}_{\SymWeak}$} & \multicolumn{1}{c?}{} \\
\multicolumn{1}{?c}{$\underbrace{\hspace{45pt}}_{\Spin^+(1,3)}$} & \multicolumn{2}{c}{$\underbrace{\hspace{100pt}}_{\tn{Weak isospin}}$} &\multicolumn{1}{c}{}& \multicolumn{1}{c?}{} \\
\multicolumn{1}{?c}{} & \multicolumn{4}{c?}{$\underbrace{\hspace{215pt}}_{\tn{Hypercharge}}$}\\
\multicolumn{1}{?c}{$\hspace{45pt}$} & \multicolumn{1}{c}{$\hspace{45pt}$} & \multicolumn{1}{c}{$\hspace{45pt}$} & \multicolumn{1}{c}{$\hspace{45pt}$} & \multicolumn{1}{c?}{$\hspace{45pt}$} \\
\toprule[1.25pt]
\end{tabular}
\end{center}
\caption{Ranges of various actions on the algebra $\SMA$. The header contains the factors of the basis elements of $\SMA$ from equation \eqref{eq:generic-element-nested-symmetries}. On the rows are represented the ranges of the action of the groups or algebras acting on the $\SMA$, that is, which factors  of the basis are affected.}
\label{tab:standard_model_sym}
\end{table}

%-----------------------------------------------------%
\section{Future plans}
\label{s:next}

Obtaining a natural relation connecting the discrete parameters and the symmetries of the Standard Model is a first step, but hopefully it may be the framework for future developments. In particular, it would be interesting if this can lead to mathematical relations between various continuous parameters like masses and coupling constants. The most natural candidate, the prediction of the Weinberg angle, was already derived, however, an explanation of why is still outside the experimental range is missing, as well as how does it explain the anomaly cancellations in the electroweak unification, if it is fixed. The number of families, the mixing matrices for neutrinos and quarks, as well as the nature of neutrinos also worth being explored within the framework of the algebra $\SMA$. The proposed model, by unifying various aspects of the Standard Model, may also be a first step toward a simpler and more insightful Lagrangian. This model clearly cannot include gravity on equal footing with the other forces, but its geometric nature and the automatic inclusion of the Dirac algebra associated with the metric may allow finding new connections with general relativity and gravity. At this stage these prospects are speculative, but this is just the beginning.
Another future step is to investigate the quantization within this framework. Since the model does not make changes to the SM, it may turn out that the Lagrangian and the quantization are almost the same as those we know. But the constraints introduced by $\SMA$ may be helpful in these directions too.
An interesting difference is the electroweak symmetry breaking induced purely by geometry, without appealing to the Higgs boson.
The Higgs boson is not forbidden by the model, being allowed to live in its usual space associated with the weak symmetry, and it is still required, at least to generate the masses of the particles. But it gained a more geometric interpretation, which may find applications in future research. The proposed model does not make any assumptions about the neutrino, except that it is represented as a $4$-spinor. This includes the possibility that it is a Weyl spinor, already refuted, or a Dirac or Majorana spinor, which is still undecided. This again depends on the dynamics. It is not excluded that subsequent development of this model may decide the problem in one way or another, at theoretical level.

%-----------------------------------------------------%
\appendix

%-----------------------------------------------------%
\section{The inner product on the ideal}
\label{s:hermitian}

To prove the relation \eqref{eq:left_ideal_hermitian_metric}, $\h(a,b)\qdvol\qvol = (a^\dagger \qvol)^\dagger b^\dagger \qvol = \qdvol a b^\dagger\qvol$ for any $a,b\in\ExtAlg\CSElectroColor$, we verify it for each element of the basis \eqref{eq:left_ideal_basis_idempotent}. We use the facts that $\ExtAlg\CSElectroColor\qvol=0$ and $\(\qdvol a b^\dagger\qvol\)^\dagger = \qdvol b a^\dagger\qvol$.

Suppose $a=1$. If $b=1$, $\qdvol a b^\dagger\qvol = \qdvol\qvol = \h(1,1)\qdvol\qvol$.
Now, suppose $b=\q{j_1\ldots j_k}$ with $k>0$. Then, $\qdvol a b^\dagger \qvol = \qdvol b^\dagger \qvol = 0$.

For $a=\q j$ it is enough to take $b^\dagger=\qd{j_1\ldots j_k}$ with $k>0$. Then, 
$\qdvol \q j b^\dagger\qvol = \qdvol \q j \qd{j_1\ldots j_k} \qvol = \qdvol (\delta_{jj_1} - \qd{j_1}\q j) \qd{j_2\ldots j_k} \qvol
=\delta_{jj_1}\qdvol \qd{j_2\ldots j_k} \qvol$. We see that the only non-vanishing case is $b=\q j$.

Now consider $a=\q {jk}$ and $b^\dagger=\qd{j_1\ldots j_k}$. The cases when $k<2$ are already checked, so it remains to check $k=2$ and $k=3$. For $k=2$,
$\qdvol \q {jk} \qd {lm} \qvol 
= \qdvol \q j (\delta_{kl} - \qd l \q k) \qd m \qvol 
= \delta_{kl} \qdvol \q j \qd m \qvol
- \qdvol \q j \qd l \q k \qd m \qvol
= \delta_{kl} \qdvol (\delta_{jm} - \qd m \q j) \qvol
- \qdvol \q j \qd l (\delta_{km} - \qd m \q k) \qvol
= \delta_{kl} \delta_{jm} \qdvol\qvol
- \delta_{km} \qdvol \q j \qd l \qvol
= \delta_{kl} \delta_{jm} \qdvol\qvol
- \delta_{km} \qdvol (\delta_{jl} - \qd l \q j) \qvol
= (\delta_{kl} \delta_{jm} - \delta_{km} \delta_{jl}) \qdvol\qvol$.
Hence, the only non-vanishing cases are $b=\pm\q j \q k$, in which case the inner product is $\pm1$, as expected. For $k=3$ it is simply to check that a factor $\q l$ remains, where $\{j,k,l\}=\{1,2,3\}$, and by multiplication with $\qvol$ gives $0$.

For $a=b=\qvol$, $\qdvol a b^\dagger \qvol = \qdvol\qvol \qdvol\qvol = \qdvol\qvol$, so the product is $1$.

All other cases are already checked, because $\(\qdvol a b^\dagger \qvol\)^\dagger=\qdvol b a^\dagger \qvol$.

%-----------------------------------------------------%
\section{From weak symmetry to \texorpdfstring{$\CCl_6$}{CCl6}}
\label{s:from_weak_to_ccl_six}

In this section I show that weak interactions lead automatically to the extension of the Dirac algebra to the complex Clifford algebra $\CCl_6$.
The Dirac algebra $\D$ is the complex Clifford algebra $\CCl_4$.
It extends to the Clifford algebra $\CCl_5$ by including the weak isospin operator $T_3$.
Then, the Dirac algebra turns out to be the even subalgebra $\CCl_5^+$ of $\CCl_5$.
The other two generators $T_1$ and $T_2$ of the weak symmetry extend the algebra $\CCl_5$ to the complex Clifford algebra $\CCl_6$.
The algebra $\CCl_5$ becomes the even subalgebra $\CCl_6^+$ of $\CCl_6$.
Hence, we have the following successive extensions, from the Dirac algebra to the complex Clifford algebra $\CCl_6$:
\begin{equation}
\label{eq:dirac_to_cl_five_to_cl_six}
\D=\CCl_4\cong\CCl_5^+\hookrightarrow\CCl_5\cong\CCl_6^+\hookrightarrow\CCl_6.
\end{equation}

From the operators $T_j$, we can construct, using the matrix $\tgamma^0=1_2\otimes\gamma^0$, where $\gamma^0=
{\scriptscriptstyle\left(\begin{array}{rr}
		0_2 & 1_2 \\
		1_2 & 0_2 \\
	\end{array}\right)}
$ in the Weyl basis, the operators
\begin{equation}
\label{eq:achiral_generators_ew}
\tau^j := 2 T_j + 2 T_j \tgamma^0.
\end{equation}
Since $\gamma^0$ swaps the left and right chiral components, we find that
\begin{equation}
\tau^j = \sigma_j\otimes 1_4.
\end{equation}

The matrices $\tau^j$ act on
${\scriptscriptstyle
\left(\begin{array}{r}
		\WeakUp \\
		\WeakDown \\
	\end{array}\right)}
	\in
	\wsu\oplus\wsd$ just like the operators $2T_j$, except that they act the same also on the right-handed components.

The matrices $\tau^j = \sigma_j\otimes 1_4$ commute with the Dirac matrices $\tgamma^\mu=1_2\otimes\gamma^\mu\in\End_\C(\wsu\oplus\wsd)$, $\mu\in\{0,1,2,3\}$. But the matrices
\begin{equation}
\omega^j := \tau^j \tgamma^5
\end{equation}
anticommute with $\tgamma^\mu$, $\mu\in\{0,1,2,3\}$, and satisfy 
\begin{equation}
\label{eq:omega_matrices}
\omega^j \omega^k = \delta_{jk} + i\epsilon^{jkl}\tau^l,
\end{equation}
for all $j,k,l\in\{1,2,3\}$.

Since $\omega^3\in\Matrix4\C\oplus\Matrix4\C\cong\CCl_5$ but is not of the form $1_2\otimes A$, $A\in\Matrix4\C$, it follows that together with matrices of the form $1_2\otimes A$, it generates the entire Clifford algebra $\CCl_5$. A Clifford basis for $\CCl_5$ is given by the matrices $\omega^3\tgamma^\mu$ and $\omega^3$. It follows that the images of the Dirac matrices in $\CCl_5$ are even elements, $\tgamma^\mu\in\CCl_5^+$.
The volume form of the Clifford algebra $\CCl_5$, which is the product of the elements of its basis, is then $(\omega^3\tgamma^0)(\omega^3\tgamma^1)(\omega^3\tgamma^2)(\omega^3\tgamma^3)\omega^3 = \tgamma^{0123} \omega^3 = -i\tau^3$.

The matrices $\omega^1$ and $\omega^2$ do not belong to $\Matrix4\C\oplus\Matrix4\C$, they belong to $\sigma_1\otimes\Matrix4\C$. They provide an extension of the algebra $\CCl_5$ to the algebra $\CCl_6$.
A Clifford basis for $\CCl_6$ is given by the matrices $\omega^1\tgamma^\mu$, $\omega^1$, and $\tau^2$.
By this, the Clifford algebra $\CCl_5$ is identified to the even subalgebra $\CCl_6^+$ of $\CCl_6$.
The volume form of the Clifford algebra $\CCl_6$, which is the product of the elements of its basis, is also $-i\tau^3$.

So far we have seen that the weak symmetry automatically leads to the successive extensions \eqref{eq:dirac_to_cl_five_to_cl_six} of the Dirac algebra to the Clifford algebra $\CCl_5$ and to $\CCl_6$,
\begin{equation*}
\D=\CCl_4\cong\CCl_5^+\hookrightarrow\CCl_5\cong\CCl_6^+\hookrightarrow\CCl_6.
\end{equation*}
Therefore, the weak symmetry requires the Dirac algebra to be extended to the complex six-dimensional Clifford algebra $\CCl_6$.

In this article I used a slightly different representation than that presented in this Section, in order to make clearer the properties of the algebra $\SMA$.

\textbf{Acknowledgments}
I wish to thank C. Castro, C. Daviau, T. Dray, C. Furey, I. Kanatchikov, A. Laszlo, G. McClellan, I. Todorov, G. Trayling, and many others, for various suggestions and feedback.

%-----------------------------------------------------%

\end{document}